\documentclass[a4paper,11pt]{article}
\pdfoutput=1 

\usepackage{jcappub} 

\usepackage{lmodern} 
\usepackage[T1]{fontenc} 

\usepackage{amsmath}
\usepackage{amssymb}

\usepackage{lineno}

\newcommand{\vecN}[1]{\mathbf{#1}} 
\renewcommand{\matrix}[1]{\boldsymbol{\mathcal{#1}}}

\usepackage{color}

\usepackage[draft]{fixme}
\usepackage{soul}
\usepackage[backgroundcolor=yellow, linecolor=black]{todonotes}

\usepackage{subfigure}

\title{\boldmath CRPropa 3 -- a Public Astrophysical Simulation Framework for Propagating Extraterrestrial Ultra-High Energy Particles}

\author[a,f]{Rafael {Alves~Batista},}
\author[a]{Andrej Dundovic,}
\author[b]{Martin Erdmann,}
\author[c]{Karl-Heinz Kampert,}
\author[b,1]{Daniel Kuempel,\note{Corresponding author.}}
\author[b]{Gero M\"{u}ller,}
\author[a]{Guenter Sigl,}
\author[a,d]{Arjen van Vliet,}
\author[b]{David Walz,}
\author[b,c,e]{Tobias Winchen}

\affiliation[a]{University of Hamburg,\\II Institut f\"ur Theoretische Physik Luruper Chaussee 149, 22761 Hamburg, Germany}
\affiliation[b]{RWTH Aachen University,\\III. Physikalisches Institut A Otto-Blumenthal-Str., 52056 Aachen, Germany}
\affiliation[c]{University of Wuppertal,\\Department of Physics, Gau\ss str.\ 20, 42097 Wuppertal, Germany}
\affiliation[d]{now at Radboud University Nijmegen, \\Department of Astrophysics/IMAPP, P.O. Box 9010, 6500 GL Nijmegen, The Netherlands}
\affiliation[e]{now at Vrije Universiteit Brussel,\\Astrophysical Institute, Pleinlaan 2, 1050 Brussels, Belgium}
\affiliation[f]{now at University of Oxford,\\Department of Physics - Astrophysics, DWB, Keble Road, OX1 3RH, Oxford, UK}

\emailAdd{rafael.alvesbatista@physics.ox.ac.uk}
\emailAdd{andrej.dundovic@desy.de}
\emailAdd{erdmann@physik.rwth-aachen.de}
\emailAdd{kampert@uni-wuppertal.de}
\emailAdd{kuempel@physik.rwth-aachen.de}
\emailAdd{gmueller@physik.rwth-aachen.de}
\emailAdd{guenter.sigl@desy.de}
\emailAdd{a.vanvliet@astro.ru.nl}
\emailAdd{walz@physik.rwth-aachen.de}
\emailAdd{tobias.winchen@vub.ac.be}

\abstract{We present the simulation framework CRPropa version 3 designed for efficient 
development of astrophysical predictions for ultra-high energy particles. 
Users can assemble modules of the most relevant propagation effects in galactic and extragalactic space, 
include their own physics modules with new features, 
and receive on output primary and secondary cosmic messengers including nuclei, neutrinos and photons.
In extension to the propagation physics contained in a previous CRPropa version, the new version 
facilitates high-performance computing and comprises new physical features such as an interface for 
galactic propagation using lensing techniques, an improved photonuclear interaction calculation, 
and propagation in time dependent environments to take into account cosmic evolution effects in anisotropy studies and variable sources. 
First applications using highlighted features are presented as well.}

\keywords{ultra high energy cosmic rays, magnetic fields, ultra high energy photons and neutrinos}

\begin{document}
\maketitle
\flushbottom

\section{Introduction}

The origin of high energy cosmic rays is still an open research question of fundamental interest for the understanding of our universe. 
Multiple aspects of cosmic rays have been investigated experimentally, most notably their steeply falling energy spectrum with a cut-off above $\sim 40$~EeV ($1~{\rm EeV}= 10^{18}\,$eV) \cite{Abraham:2010mj}. Cosmic ray arrival directions appear to be rather isotropic with a few exceptions only. As examples we mention a dipole signal above $8$~EeV~\cite{ThePierreAuger:2014nja}, and a hot spot observed in the Northern hemisphere for energies $E>57$~EeV~\cite{Abbasi:2014lda}. Concerning cosmic ray composition, measurements of the rate and shape of the depth of cosmic ray induced air showers reveal a composition with a contribution of large nuclear masses above $\sim 4$~EeV~\cite{Aab:2014aea}.

In recent years, high energy neutrinos have also been observed with a flux well above expectations from atmospheric showers \cite{Aartsen:2014gkd}. Such extraterrestrial neutrinos with energies in the PeV regime ($1~{\rm PeV}= 10^{15}\,$eV) are distributed all over the sky and may provide directional information on hadronic acceleration sites. So far, no significant autocorrelations, or correlations with matter distributions \cite{Aartsen:2014gkd}, or with cosmic ray arrival directions \cite{Aartsen:2015dml} have been observed.

Combining these experimental observations with current knowledge on large scale structures, magnetic fields and cosmic background fields leads to the following understanding of high energy cosmic rays. The lack of significant correlations of arrival directions above $\sim 1$~EeV with the galactic plane together with the limited cosmic ray confinement owing to the size of our galaxy and its magnetic field probably mean that these cosmic rays are of extragalactic origin. The exact transition from galactic to extragalactic cosmic rays is, however, not well understood yet~\cite{Giacinti:2011ww,Abreu:2012lva}. Taking advantage of the isotropic arrival distributions, at least bounds on the density of sources were derived depending on the cosmic ray energy~\cite{Abreu:2013kif}.

The observed mass composition leads to predictions for deflection of the cosmic rays in the galactic and extragalactic magnetic fields that are larger by far than expected for protons. These deflections can reach tens of degrees for iron nuclei above $\sim5$~EeV \cite{Jansson2012}, and may be one of the reasons for the rather isotropic high energy cosmic ray sky. This could also explain the lack of correlations between neutrinos and cosmic rays.

Furthermore, attempts to explain the measured energy spectrum and composition data within a one-dimensional astrophysical model prediction were undertaken where a nuclear composition has been adapted together with the slope of the injection spectrum at the sources and their maximum acceleration energy. The best fits favor injection spectra that are considerably harder than what is expected from shock acceleration theory. In addition, the preferred values for the maximally injected energies are comparatively low so that the ``cut-off'' of the spectrum may not be dominated by interactions with background fields as predicted by the so-called GZK effect~\cite{Greisen1966,Zatsepin1966} but rather be caused by the limited source energies \cite{Aab:2015bza}.

Obviously, the origin of cosmic rays is not explained easily but requires multiple aspects to be taken into account. Knowledge on many of these aspects has been acquired in the past decade or before, such that combining this knowledge in a numerical tool appears mandatory. The tool can then be used to develop different astrophysical predictions to be compared with various data distributions of different cosmic messengers. Such scenarios include models for the large scale distribution of the sources, their injection spectra and compositions, as well as models for the galactic magnetic field, and for the much less well known extragalactic magnetic fields and its structure. In this way experimental measurements can be maximally exploited to reject invalid scenarios, and to make progress in identifying an astrophysical scenario that is compatible with all measured distributions simultaneously.

The physics of nuclear decay and particle interactions with background fields is well known from laboratory experiments. Information on relevant background fields such as the cosmic microwave background (CMB) and the ultraviolet, optical, and infrared backgrounds (IRB) exists from astronomical observations, see e.g.\ \cite{Gilmore:2011ks}. Charged particle deflection in magnetic fields is precisely understood from electrodynamics. Progress in the description of the galactic magnetic field has been achieved recently by parametrizations respecting numerous Faraday rotation and starlight polarization measurements \cite{Pshirkov11,Jansson2012,Jansson2012a}. For extragalactic fields one can get information at least from simulations of the cosmological large scale galaxy structure which generate magnetic fields based on models of magneto-hydrodynamics \cite{Dolag:2008ya,Miniati:2011}. Injection spectra at sources are not yet established, but can be obtained, e.g., from shock acceleration theory.

Furthermore, at energies around $1$~EeV and below, cosmic rays propagate over cosmological distances where the cosmological expansion, variability of low energy target photon backgrounds relevant for interactions, and deflection, even diffusion, in the structured cosmic magnetic fields all become relevant. Here information results from cosmological interpretations of astronomical observations \cite{Agashe:2014kda}.

The physics of secondary messengers, namely neutrinos and photons, is also well understood. Being weakly interacting matter particles, neutrinos have cross sections small enough to be essentially unaffected by background fields and magnetic fields. High energy photon interactions are precisely described by Quantum Electrodynamics and may produce lepton pairs or even electromagnetic showers in background fields.
 
Within this context, we have developed the program CRPropa version 3 as a general and versatile simulation tool that aims at efficient development of astrophysical predictions, and produces as output primary and secondary cosmic messengers such as protons, pions, nuclei, charged leptons, neutrinos, and photons. The program is publicly available. Its modular structure allows different components of a given astrophysical scenario to be combined and assembled. Users can also extend scenarios by including their own physics modules. In comparison to the previous version CRPropa 2~\cite{Kampert2012} most of the propagation physics implemented in CRPropa 3 is identical. However, the architecture and the code implementation have been completely reworked in order to optimally profit from modern programming design and computing techniques. 

CRPropa 3 also contains new features, most notably models for the cosmological evolution of the infrared and radio backgrounds, corrections for taking into account the effects of cosmological expansion in the three-dimensional modus that is used when simulating deflections in cosmic magnetic fields, evaluation of deflections in the galactic field, updates of the implemented photodisintegration processes, and a ``four-dimensional mode'' which allows to simulate time dependent scenarios by registering particle detections within a chosen time window. Furthermore, the development and propagation of electromagnetic cascades can now be simulated numerically using a Monte Carlo approach.

The paper is structured as follows. In section \ref{Sec:InhFeat} we will briefly recap the inherited functionality and physics from the previous version CRPropa 2. Section \ref{Sec:NewFeat} explains the novel code structure of the program before introducing the main new features for galactic and extragalactic propagation. The capabilities for simulating ultra-high energy cosmic ray propagation in CRPropa 3 are demonstrated in section \ref{Sec:Examples} in a few example applications. Finally, results are summarized in section \ref{Sec:Summary}.

\section{Inherited Features from CRPropa 2}
\label{Sec:InhFeat}
In the following we will introduce the features of CRPropa version 2~\cite{Kampert2012} that have been inherited by CRPropa 3. The publicly available software tool CRPropa 2 simulates the extragalactic propagation of UHE protons, neutrons, heavier nuclei and their secondary photons, electron-positron pairs and neutrinos. Included interactions are pair production, photo-pion production and photodisintegration with the cosmic microwave background (CMB) and the UV/optical/IR background (IRB) as well as nuclear decay. It can be run in one-dimensional (1D) mode and in three-dimensional (3D) mode. In 3D mode a 3D source distribution can be specified and deflections in extragalactic magnetic fields can be simulated as well. In 1D mode the only influence of the magnetic field that is taken into account is energy loss of $e^{+}e^{-}$ pairs due to synchrotron radiation. The effects of cosmological, source and background light evolution with redshift can be included in the 1D mode as well.
For the propagation of secondary photons and $e^{+}e^{-}$ pairs a module called DINT is provided that solves the 1D transport equations for electromagnetic cascades initiated by electrons, positrons or gamma rays~\cite{Lee1998}. All these features of CRPropa 2 are available in CRPropa 3 as well.

\section{New Features in CRPropa 3}
\label{Sec:NewFeat}
\subsection{Code structure and steering}
\label{Sec:CodeStruc}
An important step has been made in redesigning CRPropa 3 following actual standards for modular codes.
Different aspects of the simulation (e.g.\ photonuclear interactions, boundary conditions, observer coordinates etc.) are separated into modules. Each module is independent of other modules and the probability, e.g.\ for an interaction, is calculated in each propagation step. To assure that the stepsize is small enough to process different modules in one propagation step in an arbitrary order, it can be limited by any of the modules. This default accuracy can be modified by the user if necessary.
Since there are no direct dependencies between modules, any combination of modules can in principle be selected, allowing for multiple use cases and to study in detail individual propagation aspects. Therefore, each module can be replaced or added, making CRPropa 3 a flexible framework that can be extended without the need to modify other components. In this context, the simulation is implemented as an user-defined sequence of independent modules, which in turn modify objects of the main cosmic ray candidate class.

The cosmic ray candidate class incorporates the relevant information about the particle propagation, for instance the actual particle type, its comoving coordinates and velocities at different times and the list of secondary particles. The candidate properties are updated at each step of its propagation until a user-provided breaking condition is satisfied.
Cosmic ray candidates can be created individually by the user or by means of a source model class, which takes the pertinent source properties, e.g.\ positions, spectrum, composition, and time-dependencies as input.

\begin{figure}[tb]
\begin{centering}
\includegraphics[scale=0.5]{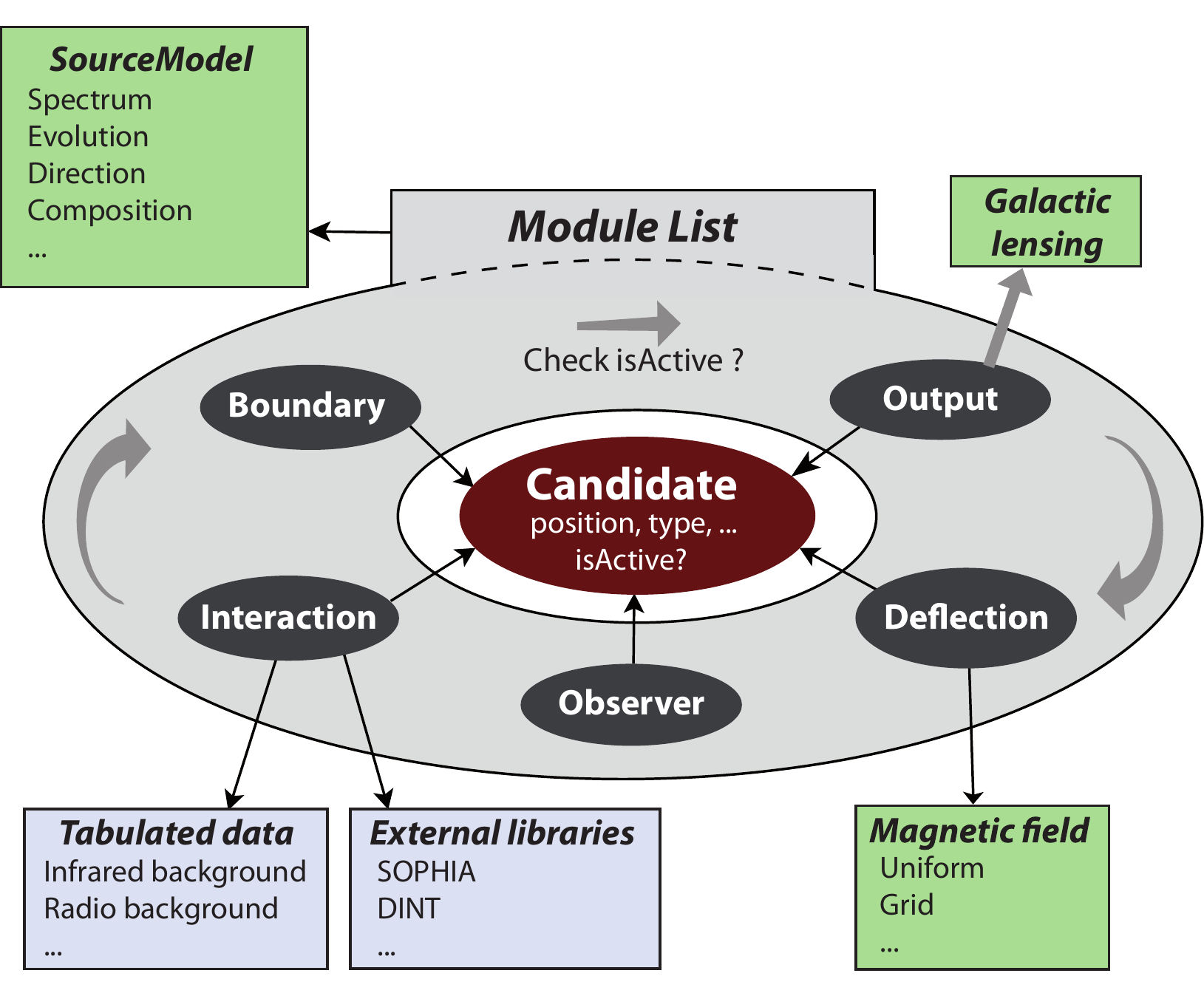}
\caption{Illustration of the CRPropa 3 modular structure. Each module contained in the module list acts on the candidate class. The \texttt{isActive} flag serves as break condition and is checked after each cycle of the module list.}
\label{fig:visu}
\end{centering}
\end{figure}

A graphical visualization of the propagation process is given in figure~\ref{fig:visu}. In this configuration, first the cosmic ray candidate is created by the source class. Then the modules sequentially process the cosmic ray candidate until the propagation is stopped by a user defined breaking condition. Output modules store all relevant information.

Cosmic ray propagation is a parallel task since interactions between cosmic rays are negligible.
Current multi-core processors can therefore be adequately utilized by just running multiple simulation instances in parallel. However, CRPropa 3 enables shared-memory multiprocessing using OpenMP\footnote{\url{http://openmp.org}} for easy multi-core simulations.
In this configuration, the ideal linear scaling is achieved up to about 8 threads. This limit is determined by critical sections in output modules and some external libraries (e.g.\ SOPHIA~\cite{Mucke2000}).

CRPropa 3 is written in C++ and interfaced to Python using SWIG\footnote{\url{http://www.swig.org}}. This allows the user to set up and steer simulations in a high level scripting language while all computations are performed with the underlying C++ code. The SWIG interface enables cross-language polymorphism, which can be used to extend a CRPropa simulation directly from the Python script that runs it.
The user can for example write a custom simulation module in Python to be used in combination with the existing C++ modules.
While the Python usage is the recommended steering mode, backward compatibility to the XML steering of CRPropa 2 is provided as well.

The performance of the code highly depends on simulation settings. For the simulation to finish, all candidates need to fulfill one of the given breaking conditions while iterating through all included modules in every propagation step. Hence, in 1D mode running time is determined solely by the execution time of the included modules, in 3D mode traversing magnetic fields dominates the overall performance: the larger deflections of cosmic rays in magnetic fields are, the smaller are the propagational steps resulting in more iterations of all modules per physical length and requiring more CPU time for the candidate to reach one of the spatial breaking conditions. To measure the performance and to find possible bottlenecks, one can use the \texttt{PerformanceModule} which calculates the percentage of time used in the total running time for every single module. More information about the performance in a specific scenario can be found in section \ref{Sec:Examples}.

\subsection{Magnetic fields}\label{sec:BFields}
As part of the restructuring, CRPropa 3 now supports any kind of magnetic field.
The only requirement is the implementation of a \texttt{getField} function in the C++ \texttt{MagneticField} interface.
This allows analytical fields, grid like fields, or fields from complex structures of any scale to be included. Table \ref{table:MagFields} summarizes the generic general and galactic magnetic field types implemented in CRPropa 3.
\begin{table}[t!]
\centering
\makebox[0pt][c]{\parbox{1.05\textwidth}{%
    \begin{minipage}[c][4.2cm][t]{0.485\hsize}\centering
    \begin{scriptsize}
        \begin{tabular}{| p{1.3cm}p{5.8cm}  | }
           \hline
           \multicolumn{2}{|c|}{\textbf{General magnetic fields}}
           \\ \hline
           \textit{Uniform}    & Magnetic field is a position independent single vector magnetic field.\\
           \textit{Grid}           & Provides a periodic magnetic field grid with trilinear interpolation, equal spacing, and different sizes along each axis.\\
           \textit{Modulated Grid} & Modulates a large scale vector field by a periodic small scale scalar field.   \\
           \textit{Turbulent}      & A random magnetic field with a turbulent spectrum~\cite{Giacinti11}.   \\
           & \\
           & \\
           & \\ \hline
        \end{tabular}
        \end{scriptsize}
    \end{minipage}
    \hfill
    \begin{minipage}[c][4.2cm][t]{0.52\hsize}\centering
    \begin{scriptsize}
        \begin{tabular}{| p{1.3cm}p{6.115cm}  | }
           \hline
           \multicolumn{2}{|c|}{\textbf{Galactic magnetic fields}}
           \\ \hline
            \textit{Toroidal Halo}     & Toroidal halo field model adopted from~\cite{Prouza:2003yf, Sun:2007mx}.\\
           \textit{Logarithmic Spiral} & Magnetic field model of axisymmetric (ASS) or bisymmetric (BSS) logarithmic spiral shape. \\
            \textit{Pshirkov 2011}          &  Pshirkov et al.\ magnetic field model, consisting of a large-scale regular (disk and halo) field~\cite{Pshirkov11}. The axisymmetric (ASS) and the bisymmetric (BSS) disk model can be chosen. \\
           \textit{JF 2012}   & Implementation of the Jansson \& Farrar magnetic field model, consisting of a large-scale regular and random (striated) field and a small-scale random (turbulent) field~\cite{Jansson2012, Jansson2012a}.   \\  \hline
        \end{tabular}
        \end{scriptsize}
    \end{minipage}
    \hfill
        \caption{Implemented general and galactic magnetic field types.}
    \label{table:MagFields}
}}
\end{table}

In addition, CRPropa 3 utilizes third party libraries to access data from adaptive mesh refinement and smooth particle simulations:
The SAGA (SQLite AMR Grid Application) code\footnote{\url{https://github.com/rafaelab/saga}} uses R-Trees for fast access to RAMSES \cite{Teyssier:2006us,Fromang:2006aw} magnetohydrodynamical simulations data.
Quimby\footnote{\url{https://forge.physik.rwth-aachen.de/projects/quimby}}~\cite{Muller:2015gkr} is a multi-resolution grid library for fast access to huge compressed magnetic field grids and provides access to smooth particles from the GADGET~\cite{Springel:2000yr,Springel:2005mi} file format.

The new structure also allows the implementation of magnetic field decorators, which can modify the given magnetic field on the fly.
A periodic decorator turns any magnetic field into a periodic one, with the option of reflective boundaries.
An evolution decorator allows for a cosmological scaling of type $B_{\rm com}(z) = B(0) (1+z)^{m}$, where $B_{\rm com}$ is the comoving magnetic field and $m$ the magnetohydrodynamic amplification of the field, cf.\ section \ref{Sec.CosEffects}.
Multiple magnetic fields can also be grouped together and their magnetic fields can be superpositioned (added) in a \textit{list}.

\subsection{Galactic propagation}\label{sec:GalProp}
The propagation framework described above can, in principle, model cosmic ray
propagation on any scale. When modeling the propagation of extragalactic cosmic
rays entering the Milky Way with CRPropa 3 interactions can be neglected and
only deflections in the magnetic field need to be considered. Instead of
loading tabulated magnetic field data using the existing modules, dedicated
modules describing the deflection in specific magnetic field models are used.
The galactic magnetic field models proposed by Jansson \&
Farrar~\cite{Jansson2012, Jansson2012a} and Pshirkov et al.~\cite{Pshirkov11} are implemented and can be used as examples for other
models. Forward and backward tracking of particles can be achieved by injection
of the corresponding particles into the simulation.

However, to account for galactic magnetic field effects over large distances
using forward propagation is computationally inefficient as most of the
simulated particles miss the observer.  As alternative, backtracking of cosmic
rays with opposite charge from the observer to the edge of the galaxy is a much
more efficient option to study possible trajectories of cosmic rays inside the
Milky Way.  In CRPropa 3 we provide an interface to the `lensing technique'
developed for the PARSEC software~\cite{Bretz2014} that allows an efficient
usage of backtracking simulations to account for the effects of deflection in
the galactic magnetic field in forward simulations of extragalactic cosmic
rays.

In the lensing approach, the trajectories of backtracked particles with
rigidity $E_i/Z_i$ are stored in matrices $\matrix{L}_i$ which are used to map
discrete directions (pixel)  indexed with $n$ outside the galaxy to discrete
observed directions indexed with $m$ on Earth. The set of matrices $\lbrace
\matrix{L}_1 \cdots \matrix{L}_N \rbrace$ form the `galactic lens' that
completely describes the deflection of extragalactic cosmic rays in the galactic
magnetic field model. Using the matrices, a vector $\vecN{p}_{\rm eg}^i$ of the
probabilities to observe a cosmic ray at energy $E_i$ from direction $n$ at the
edge of the galaxy can be transformed by a matrix vector multiplication
\begin{equation}
	 \matrix{L}_i \cdot \vecN{p}_{\rm eg}^i = \vecN{p}_{\rm obs}^i
	\label{eq:GalacticLensesEquation}
\end{equation}
to obtain  the probability distribution $\vecN{p}_{\rm obs}^i$ for energy
$E_i$.  To avoid artificial distortions of the energy spectrum in this
approach, all matrices have to be normalized by the maximum of unity norms
$\Vert \matrix{L}_i \Vert_1$ of all matrices in the set.

The observed probability distributions can either be analyzed directly or
be used to generate sets of individual simulated cosmic rays. Convenient interfaces to create the probability
distributions from extragalactic CRPropa simulations and to sample data from
the probability distributions are included in CRPropa 3.

The observed and injected directions are
discretized using the HEALPix scheme~\cite{Gorski2005}. The matrices
are accessed in sparse compressed column major format using the
Eigen\footnote{\url{http://eigen.tuxfamily.org}} template library for
linear algebra to minimize memory consumption and disk storage space while
maximizing the computational performance for the matrices-vector
multiplications.

\subsection{Photonuclear interactions}
\label{sec:Interactions}
Photonuclear interactions with the CMB and IRB are the dominant energy loss processes for ultra-high energy protons and nuclei.
While the spectral shape of the CMB is well known, various models exist for the IRB.
In addition to the Kneiske~2004~\cite{Kneiske2004} model that was considered in CRPropa 2, the following IRB models are now available as well: Stecker~2005~\cite{Stecker2005}, Franceschini~2008~\cite{Franceschini2008}, Finke~2010~\cite{Finke10},  Dominguez~2011~\cite{Dominguez11} and Gilmore~2012~\cite{Gilmore:2011ks}.

Furthermore, the implementation of photonuclear interactions has been improved in several ways.
The pion production cross sections are now considered over a wider range of photon energies, and the integration procedure for calculating the interaction rates as well as the interpolation of the tabulated interaction rates were enhanced.
This addresses the suggestions by Kalashev and Kido~\cite{Kalashev2014}.
Additionally, the following improvements regarding photodisintegration were implemented.
In CRPropa 2 photodisintegration cross sections were mainly obtained with TALYS 1.0~\cite{Goriely2008}; see ref.~\cite{Kampert2012} for a detailed description.
These cross sections are updated using the more recent TALYS version 1.6 (cf.\ ref.~\cite{TALYSmanual} for a full description of the changes in TALYS).
More importantly, the parameters used in TALYS to model the giant dipole resonance have been adjusted to match the results given in \cite{Khan:2004nd}.
The complete list of used parameters is given in appendix~\ref{appendix:GDR}, table \ref{tab:GDRparameters}.
We verified that the recently published TALYS version 1.8 gives the same results with the used settings.

In CRPropa~3 the photonuclear cross sections are tabulated for photon energies in the rest frame of the nucleus of $\epsilon' = 0.2 - 200$\,MeV in logarithmic steps of $\Delta \log_{10}(\epsilon'/\mathrm{MeV}) = 0.01$.
This is done for all 169 isotopes in the range $A = 12 - 56$, $Z \leq 26$ with a lifetime of $\tau > 2\,$s.
The branching ratios are taken into account for every channel that is computed in TALYS, namely with a simultaneous emission of up to eight nucleons in form of protons, neutrons, $d$, $t$, $^3$He and $^4$He nuclei.
In practice, a large fraction of the resulting more than 25000 channels is of negligible impact for cosmic ray propagation.
Thus, to increase performance, channels with branching ratios of less than 5\% at every energy in the tabulated range are neglected, and the branching ratios of the remaining channels are scaled up accordingly.
For the total cross section, however, all channels are considered.
In contrast to the thinning procedure adopted in CRPropa~2, this procedure prevents a small systematic overestimation of the mean free path.
For isotopes with mass numbers $A < 12$, the same set of cross sections as in CRPropa~2 is used, with the exception of $^6$Li, which is now modeled using the parametrization from Kossov~\cite{Kossov2002}; as well as omitting $^6$He and $^9$Li, which, due to their short lifetime $\tau < 2s$, exhibit negligible photodisintegration.
As a result, TALYS is not used for any isotope with $A < 12$, as is recommended in ref.~\cite{TALYSmanual}.
In total, photodisintegration is considered for 183 isotopes and 2200 channels.

Alternatively, CRPropa 3 provides the option of modeling photodisintegration for all isotopes using the parametrization from Kossov~\cite{Kossov2002}.
While the Kossov parametrization models both photodisintegration and photo-hadronic interactions, only the photodisintegration part up to the pion production threshold is considered here, since pion production is treated separately in CRPropa.
Also, since the parametrization only models the total cross section, branching ratios are computed with TALYS as described above.
A comparison of the two approaches to the available measured nuclear cross sections from the IAEA handbook on photonuclear data \cite{iaea-tecdoc-1178} indicates a slightly better agreement with the TALYS version.
The differences in typical UHECR scenarios is at level of 10\% in the spectral flux; see ref.~\cite{SimProp_CRPropa} for more details.

\subsection{Cosmological effects}
\label{Sec.CosEffects}
Cosmological effects can affect the propagation of UHECRs. The adiabatic expansion of the universe implies a reduction in the momentum of a cosmic ray by a factor $(1+z)^{-1}$, where $z$ is the redshift whose evolution is given by
\begin{equation}
 \left| \frac{{\rm d}t}{{\rm d}z} \right| = \frac{1}{H_0(1+z) \sqrt{\Omega_m(1+z)^3+\Omega_\Lambda}}
\end{equation}
in the standard $\Lambda$CDM cosmology. Here, the Hubble parameter at present time is $H_0\approx67.3$ $\mathrm{km/s/Mpc}$, $\Omega_m\approx0.315$ is the density of matter in the universe, encompassing both baryonic and dark matter, and $\Omega_\Lambda\approx0.685$ is the cosmological constant, assuming a flat universe ($\Omega_{\rm tot} = \Omega_{m} + \Omega_{\Lambda} =1  $)~\cite{Agashe:2014kda}.
 
Interaction rates for different processes depend on the redshift at which the cosmic ray interacts with the photon backgrounds, because the number density and spectral shape of these radiation fields are evolving with time.
In the case of the CMB, the spectral number density evolves passively as
$n(\epsilon,z) = (1+z)^2 \, n(\epsilon/(1+z), z=0)$.
On the other hand, the IRB is determined by the sum of galactic luminosities integrated over the entire age of the universe, hence its dependence is non-trivial. Similarly to CRPropa 2, we consider the evolution of the IRB through a global scaling factor $s(z)$, thereby assuming the spectral shape of the number density to be constant.
The scaling factor is obtained for each IRB model (cf.\ section \ref{sec:Interactions}) as the integral over the comoving spectral number density, relative to its value at the present time.
Using this approximation, the interaction rate is given by
\begin{align}
  \lambda^{-1}(E,z) &= s(z) (1+z)^3 \lambda^{-1}(E(1+z), z=0) \\
  s(z) &= \begin{cases} \quad\quad 1 & \text{CMB} \\ \frac{\int n(\epsilon,z) {\rm d}\epsilon}{\int n(\epsilon,0) {\rm d}\epsilon} & \text{ IRB} \end{cases}
\end{align}
In contrast to CRPropa 2, the integral for the IRB is performed over the entire modeled energy range, instead of just the part where the IRB dominates over the CMB.
We have also tested a more detailed treatment, in which the exact redshift evolution of the IRB is taken into account. We found that this affects the propagated spectra due to the photopion production by less than 1\% in realistic source scenarios such as the one shown in sec. \ref{sec:Benchmark_example}. In the more extreme case of high redshift sources ($z\geq 2$) and only photopion production in the IRB, this value is below 10\%. This uncertainty for photodisintegration is expected to be roughly the same, although this case was not explicitly tested.

Including cosmological effects in the simulation requires an {\it a priori} knowledge of the propagation length, thus the redshift at the time of emission. In a one-dimensional environment this is trivial, since the redshift corresponds to the distance of the source to the observer. In a three-dimensional environment, however, the situation is more complicated because the effective propagation length can change due to deflections caused by intervening magnetic fields, and due to the redshift dependence of the photon backgrounds. To take into account simultaneously cosmological effects as well as deflections by magnetic fields, in CRPropa 3 a generalization of the 3D tracking of particles encompassing the time, respectively redshift, dimension is introduced, which is henceforth called `4D mode'. This new feature extends the notion of a three-dimensional observer to four dimensions, so that the observer is considered a hypervolume composed by a sphere of a given radius $R_{\rm obs}$ and a redshift window of size $\Delta z_{\rm obs}$ around $z=0$. The calculations for $z<0$ is obtained by extrapolating the corresponding quantities down to the desired value.

The redshift window ($\Delta z_{\rm obs}$, corresponding to $\Delta t_{\rm obs}$) has to be smaller than all relevant time scales on which the cosmic flux could vary, unless one is interested in studies of time variability of sources, in which case the window should be as small as possible. This means that if the Hubble time is expressed by $t_H$, then $\Delta t_{\rm obs} / t_H \sim 0.1$ is already a sufficient constraint if the density of background photons vary slowly with redshift. Moreover, at the extreme high energy part of the spectrum ($E \gtrsim 10 \text{ EeV}$) we are limited by interaction horizons, which are smaller than $\lesssim 100 \text{ Mpc}$, corresponding to a small redshift ($z \lesssim 0.02$). For $E\lesssim 1 \text{ EeV}$ adiabatic losses are much larger than the others, and that is the region where the effects of the redshift window would matter. We have performed tests to check the convergence of the spectrum for different $\Delta z_{\rm obs}$ in a scenario with null magnetic field. The differences between $\Delta z_{\rm obs}=0.05$ and $\Delta z_{\rm obs}=0.20$ are of the order of 10$^{-5}$ at 1 EeV and 10$^{-6}$ at 0.1 EeV.

CRPropa 3 uses comoving coordinates internally. This implies that the flux dilution of comoving magnetic fields with redshift $\propto (1+z)^2$ is implicitly taken into account. Additionally, a redshift scaling of the form $B_{\rm com}(z) = B(0) (1+z)^m$ can be applied to any magnetic field to account for a general field damping resulting in a total magnetic field evolution of $B(z) = B_{\rm com}(z)(1+z)^2$. An explicit evolution of structured extragalactic magnetic field models resulting from magnetohydrodynamical simulations is currently not implemented, but can be added as described in section \ref{sec:BFields}.

The use of comoving coordinates also implies that the comoving density of an arbitrary distribution of sources is kept constant. The redshift at the time of emission can be set by hand, or randomly picked according to a source evolution of type $f(z) \propto (1+z)^{m}$, where e.g. the star formation rate for $z < 1$ is typically described with $m=3.4$~\cite{Hopkins:2006bw}. Note, that the evolution parameter $m$ for the source density and magnetic field need not be the same, and are completely unrelated.

\subsection{Secondary messengers}
Neutrinos and gamma rays produced in UHECR propagation are important
messengers as well. Neutrinos result from the decay of the charged pions
produced in photo-pion production and in nuclear decays. Being weakly
interacting particles they are only subject to adiabatic energy loss and
propagate on straight lines. Compared to the previous versions of CRPropa
neutrinos are now processed directly by the module chain, allowing an
analysis of secondary neutrinos by using the same observer objects as for the
UHECR nuclei in a single simulation chain.

High energy gamma rays are produced in the decay of neutral pions and by
inverse Compton scattering of high energy electrons with background photons.
Secondary photons from photonuclear interactions that result from secondary nuclei decaying from excites states are not considered for now~\cite{Anchordoqui:2006pd}. Electrons are created in electron-positron production, beta decay of nuclei, and the decay of muons from the decay of charged pions. Together they form an
electromagnetic cascade, in which the high energy photons dominantly interact with background photons $\gamma_b$ via pair
production $\gamma\gamma_b \rightarrow e^+e^-$ and double pair production
$\gamma\gamma_b \rightarrow e^+e^-e^+e^-$. The electrons lose energy via
inverse Compton scattering $e^\pm\gamma_b \rightarrow e^\pm\gamma$, triplet pair
production $e^\pm\gamma_b\rightarrow e^\pm e^+e^-$ and synchrotron radiation
from gyrating in magnetic fields.

To calculate the development of the electromagnetic cascade CRPropa 3 provides
interfaces to shipped versions of the specialized codes DINT~\cite{Lee1998}, which was also
part of the previous versions of CRPropa, and as a new feature also to
EleCa~\cite{Settimo2013}. DINT calculates the observed spectra based on the transport equations given in ref.~\cite{Lee1998}. It is
therefore computationally efficient and thus particularly suited for calculations
at lower energies where the cascade consists of many particles. Conversely,
EleCa provides a full Monte Carlo tracking of the individual particles with
stochastic treatment of the energy losses. Both codes currently focus on
1D propagation allowing the investigation of the resulting energy spectra of secondary particles.

The shipped version of the EleCa code has been improved regarding its
performance resulting in a speed up of a factor of 3.6 compared to the baseline in a
benchmark application. The speed up was achieved by replacing the on-the-fly
calculation of the differential cross sections for the individual processes by
interpolation between precalculated values and by code optimizations.  The
relative difference in the differential cross section between the
on-the-fly calculation and the interpolation is smaller than $10^{-8}$ and thus
negligible.

To access the cascade codes, the secondary photons and electrons generated in
ultra-high energy nuclei propagation are written to a separate file which can then be
further processed outside of the module chain by DINT or EleCa.
Additionally, we provide an interface to a combined propagation in which
particles above a customizable threshold energy are propagated individually
with EleCa, while particles below the threshold energy are processed with
DINT, cf.\ section \ref{Sec.AppSecondary}.

 \section{Example Applications Focussing on New Features}
\label{Sec:Examples}
\subsection{1D simulation including secondary particles}
\label{Sec.AppSecondary}
To demonstrate the new features of CRPropa 3 we investigate the production of secondary
messengers generated in the propagation of UHECR proton and iron primaries in a
1D simulation. We inject 10000 primaries from homogeneously distributed
sources from a minimum distance of 3~Mpc up to a maximum
distance of 1000~Mpc. The energies of the primaries are
selected with a spectrum following a power law with index $\gamma = -1$ between
a minimum energy of 1~EeV and a maximum energy of 1000~EeV. We
include all available energy losses as described in the previous section using
the default photon background models in the simulation.

\begin{figure}[tb]
	\includegraphics[width=.49\textwidth]{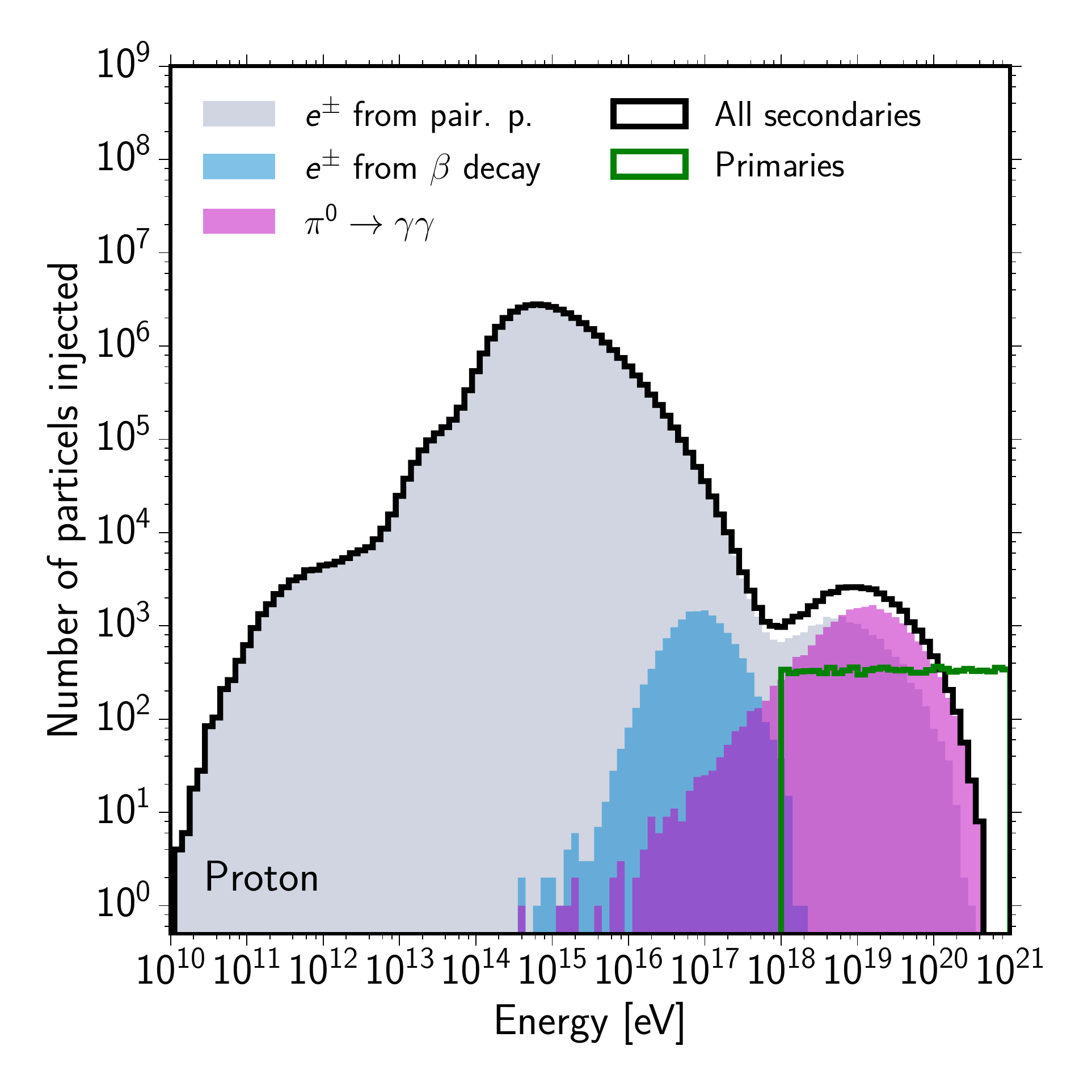}
	\includegraphics[width=.49\textwidth]{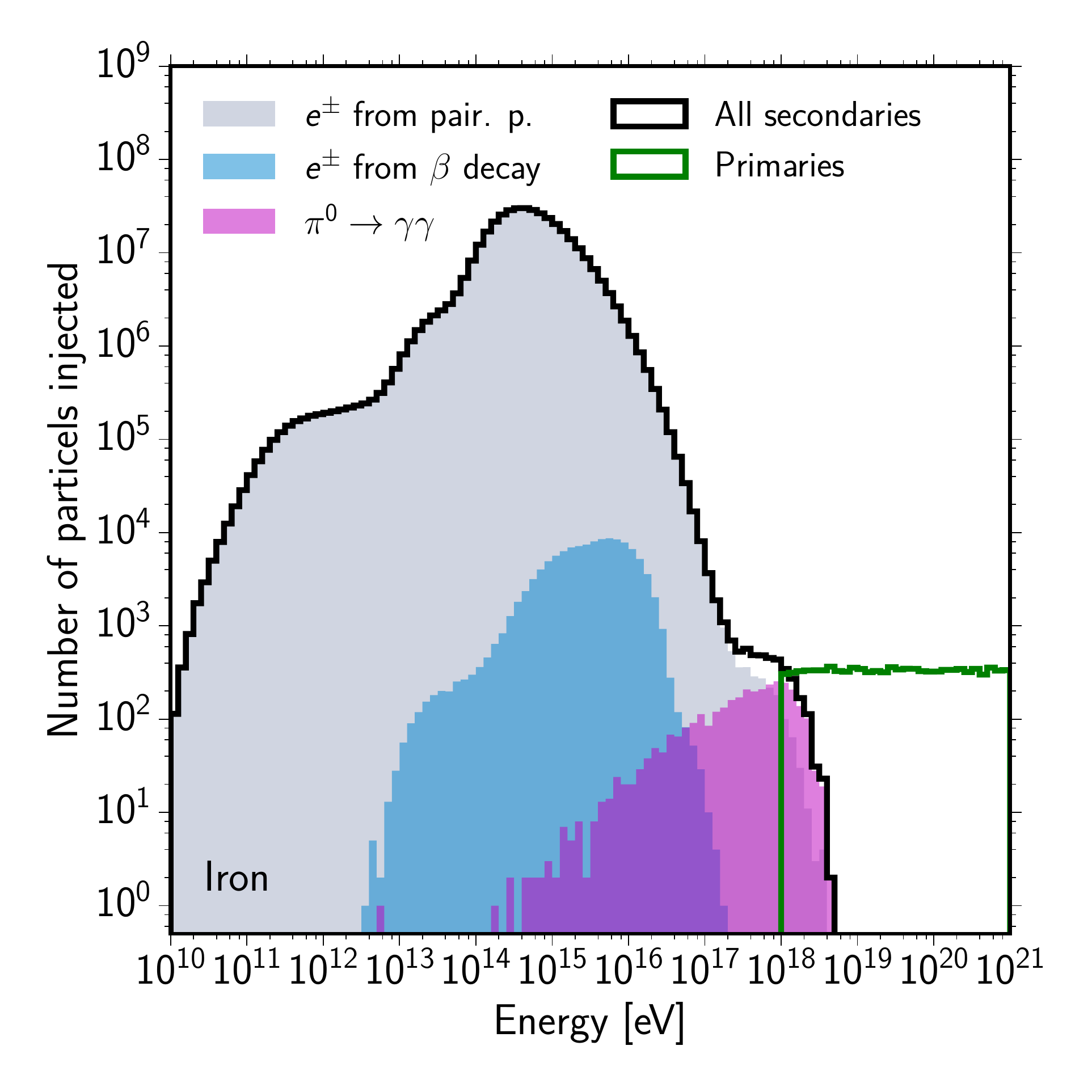}
	\caption{Histograms of the number of secondaries injected into the
	electromagnetic cascade from the propagation of proton (left) and iron
(right) primaries for the individual creation processes integrated over the
whole propagation distance, see section \ref{Sec.AppSecondary} for details. As comparison, the injected primary spectrum is shown as well (green). The observed photon and neutrino fluxes from this simulation are shown in figures \ref{fig:observed_photons} and \ref{fig:observed_neutrino_flux}.}
	\label{fig:injected_secondaries}
\end{figure}

The energy distributions of the secondaries injected into the electromagnetic
cascade are shown in figure \ref{fig:injected_secondaries} separated into
generating processes and type of primary.
For protons, the majority of secondary particles are low energy electrons
created in electron pair production. Higher energy electrons are also generated
in the $\beta$-decay of neutrons, which were created in photo-meson production.
The highest energetic secondaries are photons from the decay of the neutral
pions.  For iron primaries, the created secondaries are also electrons from
pair production and high energy photons from the decay of neutral pions.
However, the number of photons is lower and the photons have a lower mean
energy due to the lower energy per nucleon of the particles. The number of
created electrons from pair production, however, is higher than in the proton
case due to the larger number of secondary nuclei that are created from
photodisintegration. No secondaries from $\beta$-decay of radioactive nuclei
are generated in this simulation.

\begin{figure}[tb]
	\includegraphics[width=.49\textwidth]{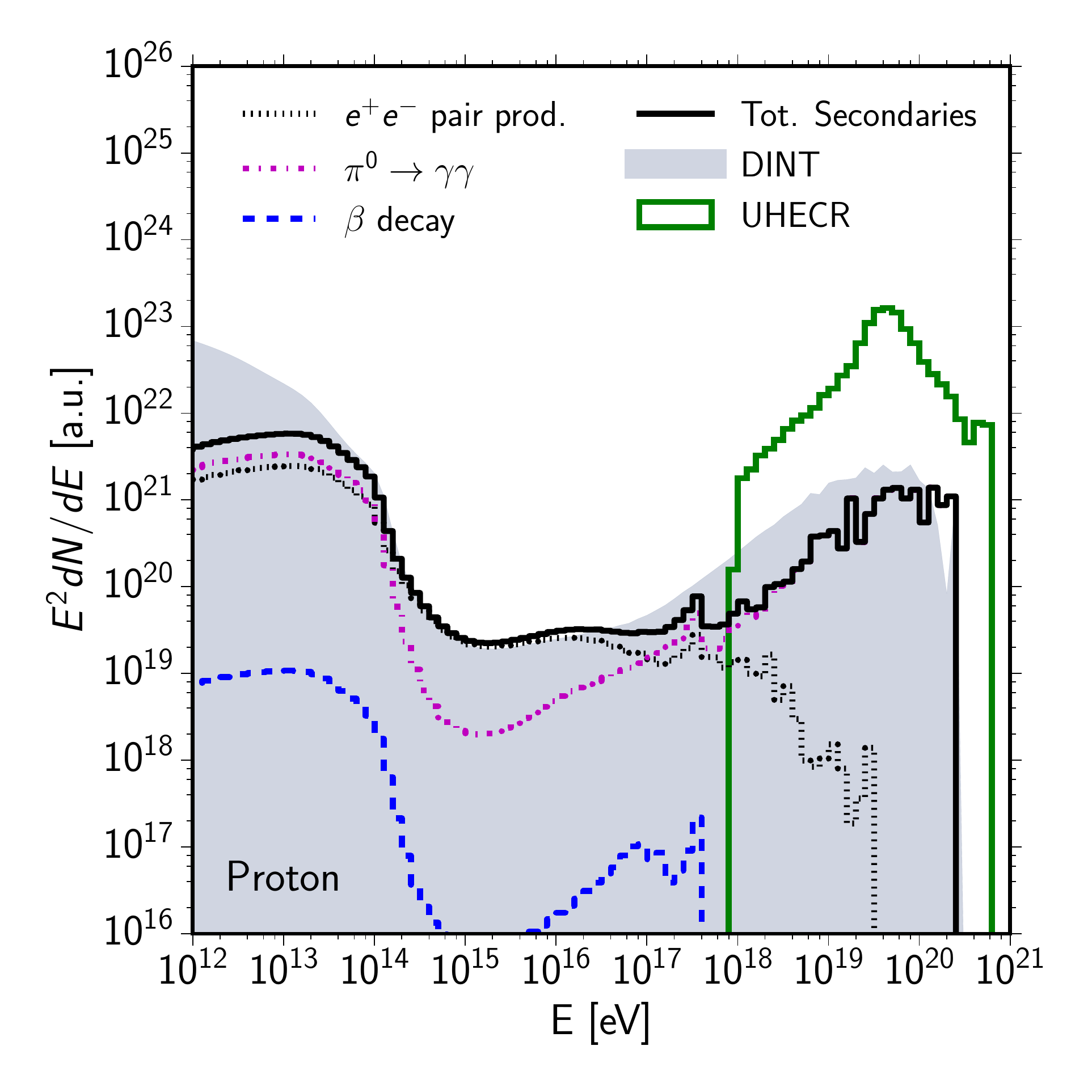}
	\includegraphics[width=.49\textwidth]{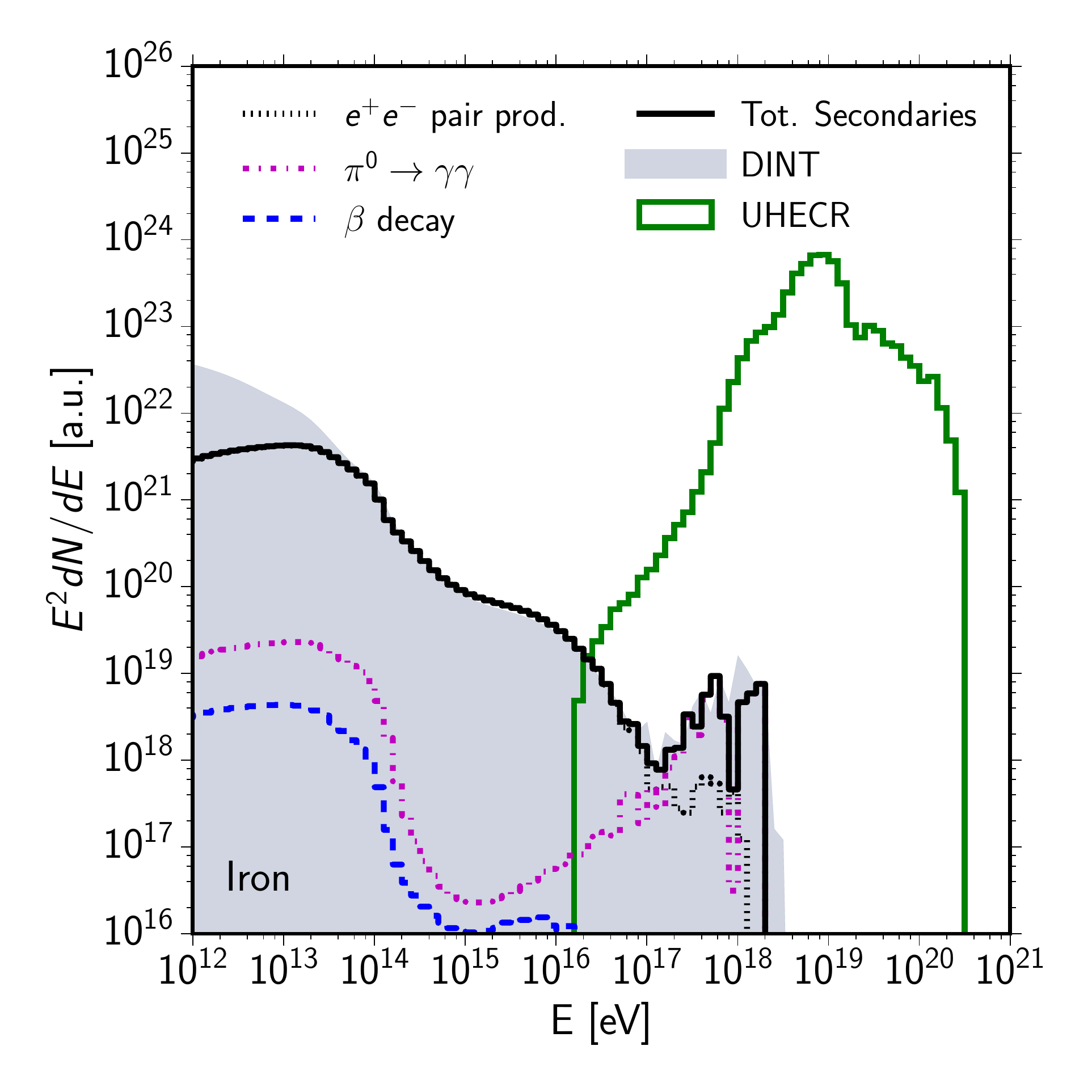}
	\caption{Histograms of the observed photon flux from proton (left)
	and iron (right) primaries separated by individual processes for the combined
propagation. As comparison, the DINT only propagation as in previous CRPropa
versions is shown as a gray background, and the propagated UHECR spectrum is shown as a green histogram.}
	\label{fig:observed_photons}
\end{figure}

The secondary photons and electrons are then propagated with DINT and the
combined DINT-EleCa propagation module using a threshold energy of 0.5~EeV. The observed spectra are shown in
figure~\ref{fig:observed_photons} as histograms for the combined
propagation and gray surface for the DINT only propagation.
For energies above a few hundred TeV ($1~{\rm TeV}= 10^{12}\,$eV) the results of both propagation
modules are in reasonable agreement for iron primaries, considering the relative low number of high energy photons in this simulation. 
For energies below a few hundred TeV the propagation with EleCa for the highest energetic photons and electrons results in a decreased spectrum for proton and iron primaries compared to the DINT only propagation. Furthermore, in case of proton primaries also differences between both modules above a few hundred PeV are visible.

The inclusion of secondary electrons increases the size of
the simulation output significantly, amounting to 2.5~GB of
uncompressed output in ascii format in case of the proton primaries and 23~GB
for the iron nuclei. The secondary electrons are
important for the estimation of the photon flux below
approximately 5~EeV.

To benefit from the new modular and flexible structure we are currently developing a photon and electron propagation module for simulating electromagnetic cascades which will be integral part of the CRPropa 3 framework.  The results will be reported separately. 

\begin{figure}[tb]
	\includegraphics[width=.49\textwidth]{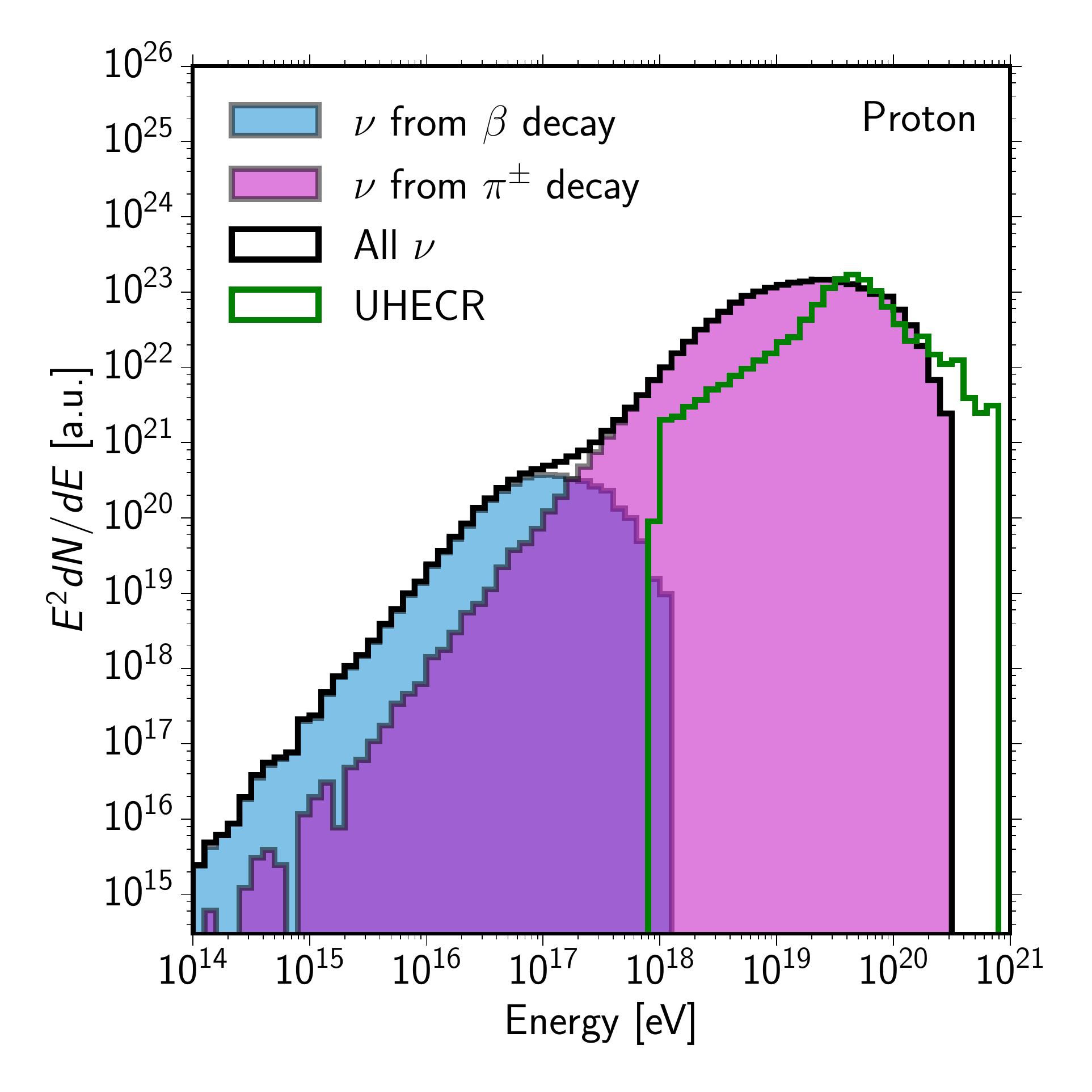}
	\includegraphics[width=.49\textwidth]{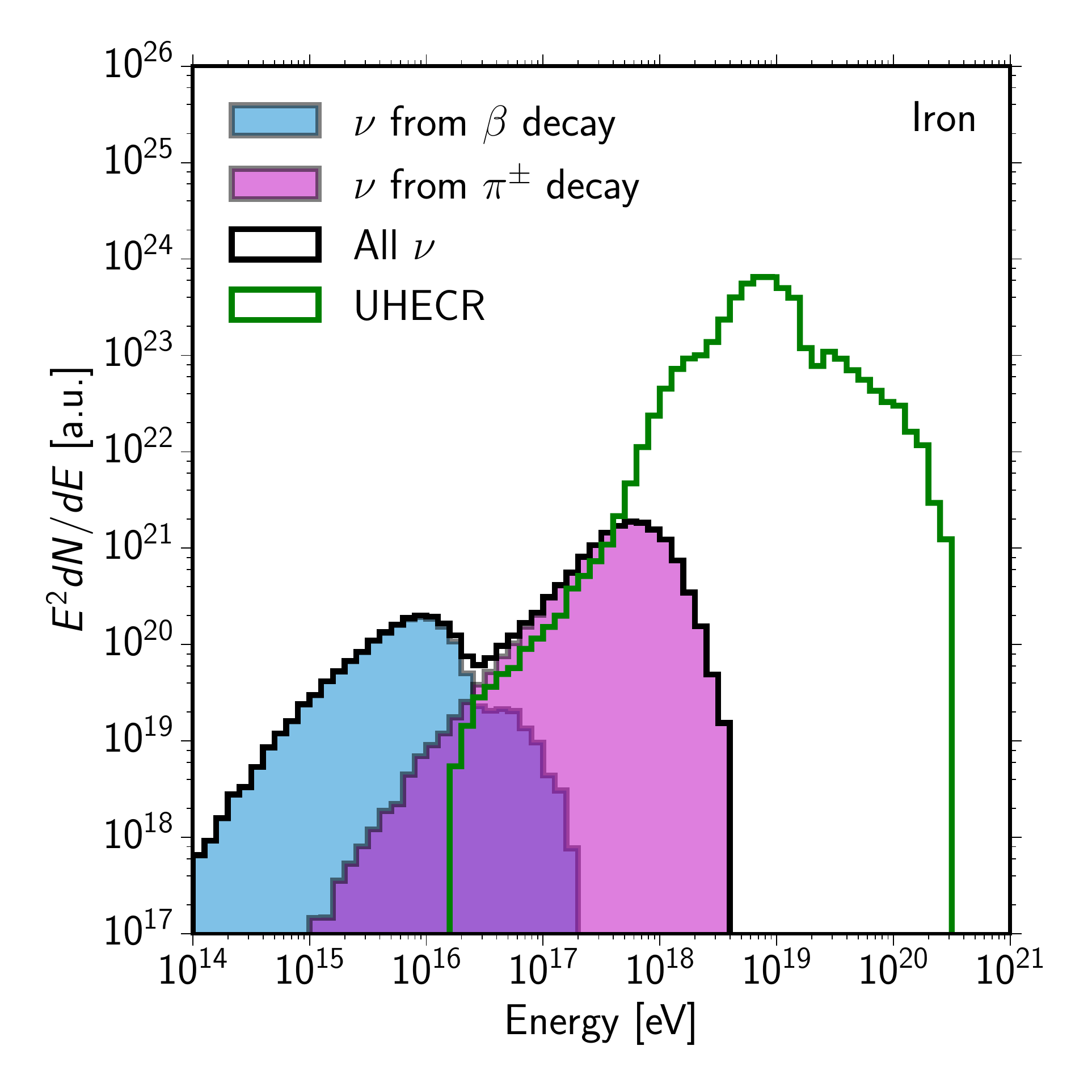}
	\caption{Histograms of the observed neutrino flux from proton (left) and iron (right) primaries separated by individual creation processes. As comparison, the propagated UHECR spectrum is show as a green histogram. }
	\label{fig:observed_neutrino_flux}
\end{figure}

The resulting flux of secondary neutrinos is displayed in
figure~\ref{fig:observed_neutrino_flux}. In the case of proton primaries,
approximately four times more neutrinos are generated in decays of $\pi^\pm$
compared to $\beta$-decays. In the case of iron primaries, approximately ten times
more neutrinos are generated from $\beta$ decay than from the decay of
$\pi^\pm$.  For the same number of injected particles, approximately twice the
number of neutrinos are observed for iron primaries than for proton primaries.
However, in this case twenty times more UHECR are observed from the same number of injected primaries.

\subsection{3D simulation including extragalactic and galactic deflections}
\label{sec:Benchmark_example}%

To show the capabilities of the 3D mode of CRPropa 3 an example is presented here including deflections in extragalactic and galactic magnetic fields as well as a source distribution following the large-scale matter density. In this scenario the sources are distributed randomly with a source density of $10^{-3}~\text{Mpc}^{-3}$ following the large-scale structure formation simulations from Dolag et al.~\cite{Dolag2005}, which have been constrained by the observed large-scale density field with a radius of 110~Mpc around the Milky Way.

The structure of the implemented extragalactic magnetic field model in this example is obtained from the same constrained simulations by Dolag et al. The overall magnetic field strength, however, is smaller in their simulations than in the similar but unconstrained simulations done by Sigl et al.~\cite{Miniati2004}. Sigl et al.\ scaled the magnetic field strength such that the magnetic field in the core region of a Coma-like galaxy cluster is of the same order as indicated by Faraday rotation measures. To emphasize cosmic-ray deflections, we derived the magnetic field strength for this example from the relation between matter density and magnetic field strength obtained from the simulations by Sigl et al. This has been done by assigning to each cell in the large-scale matter density grid from Dolag et al.\ the magnetic field strength from the simulations by Sigl et al.\ corresponding to the matter density in that cell. One thus obtains a magnetic field modulation grid (see also section~\ref{sec:BFields}) of $256^3$ cells covering a volume of $\sim (132\text{ Mpc})^3$. A higher-resolution magnetic field with Fourier modes taken from a Gaussian distribution with $\langle |B({\bf k})|^2\rangle$ given by a Kolomogov power spectrum and random polarization, with a coherence length of 500~kpc and a total volume of $\sim (13.2\text{ Mpc})^3$, is then periodically repeated to cover the complete modulation grid. Cosmic rays with trajectory lengths up to 4~Gpc are taken into account by employing reflective boundary conditions when they reach the edge of the simulation box.

Apart from magnetic field deflections, all available photonuclear and decay interactions (see section~\ref{sec:Interactions}) on both the CMB and the IRB have been included in the extragalactic propagation as well. In this 3D simulation the photon background is taken as time-independent and adiabatic losses are neglected. The implemented IRB model is that from Gilmore 2012~\cite{Gilmore:2011ks}. Two different scenarios are presented here, one for pure proton injection at the sources and one for pure iron injection. In both cases the cosmic rays are initiated at their sources following a power law spectrum with a broken exponential cutoff:
\begin{equation}\label{eq:PowerLawInjection}
	\frac{\text{d}N}{\text{d}E} \propto \left\{
	\begin{array}{l l}
 	(E/E_0)^{\gamma} & E \leq E_{\text{cut}} \\
 	(E/E_0)^{\gamma}\exp(1-E/E_{\text{cut}}) & E > E_{\text{cut}}
	\end{array}
	\right.
\end{equation}
with $N$ the number of injected particles and $E$ the particle energy. For this example the particles are injected with a spectral index of $\gamma = -1.5$ and cutoff energy $E_{\text{cut}} = 780$~EeV down to a minimum energy of $E_0 = 1$~EeV.

We consider the observer to be a sphere with a given radius $R_{\rm obs}$, which should be small enough to avoid spurious effects arising from fluctuations in the magnetic fields in different regions of the observer sphere. The smaller $R_{\rm obs}$ is, the closer the simulated observer is to reality, but the larger the required computational time for a sufficient number of observed events. A good trade-off is to take $R_{\rm obs} \lesssim \ell_s$, where $\ell_s$  is the typical scattering length for a charged particle in the magnetic field. This condition ensures that no significant diffusion will occur within the observer. From numerical studies of the trajectories of protons in this magnetic field we have obtained $\ell_s \sim 300 \text{ kpc}$. For this reason we set $R_{\rm obs} = 100 \text{ kpc}$.

Due to the finite size of the observer, another important factor that should be taken into account is the multiplicity of detections. In CRPropa it is possible to consider multiple detections of the same particle. However, in reality this would not happen and simulating multiple detections of the same cosmic ray is not in accordance with Liouville's theorem. To solve this issue we randomly select one hit among all detections of the same particle. Therefore, when multiple hits within the same periodic box from the same original cosmic ray are registered, one of these hits is chosen randomly and only that hit is used in the analysis.

\begin{figure}[tb]
	\subfigure[3D proton, no galactic magnetic field]{
    	\includegraphics[width=.49\textwidth]{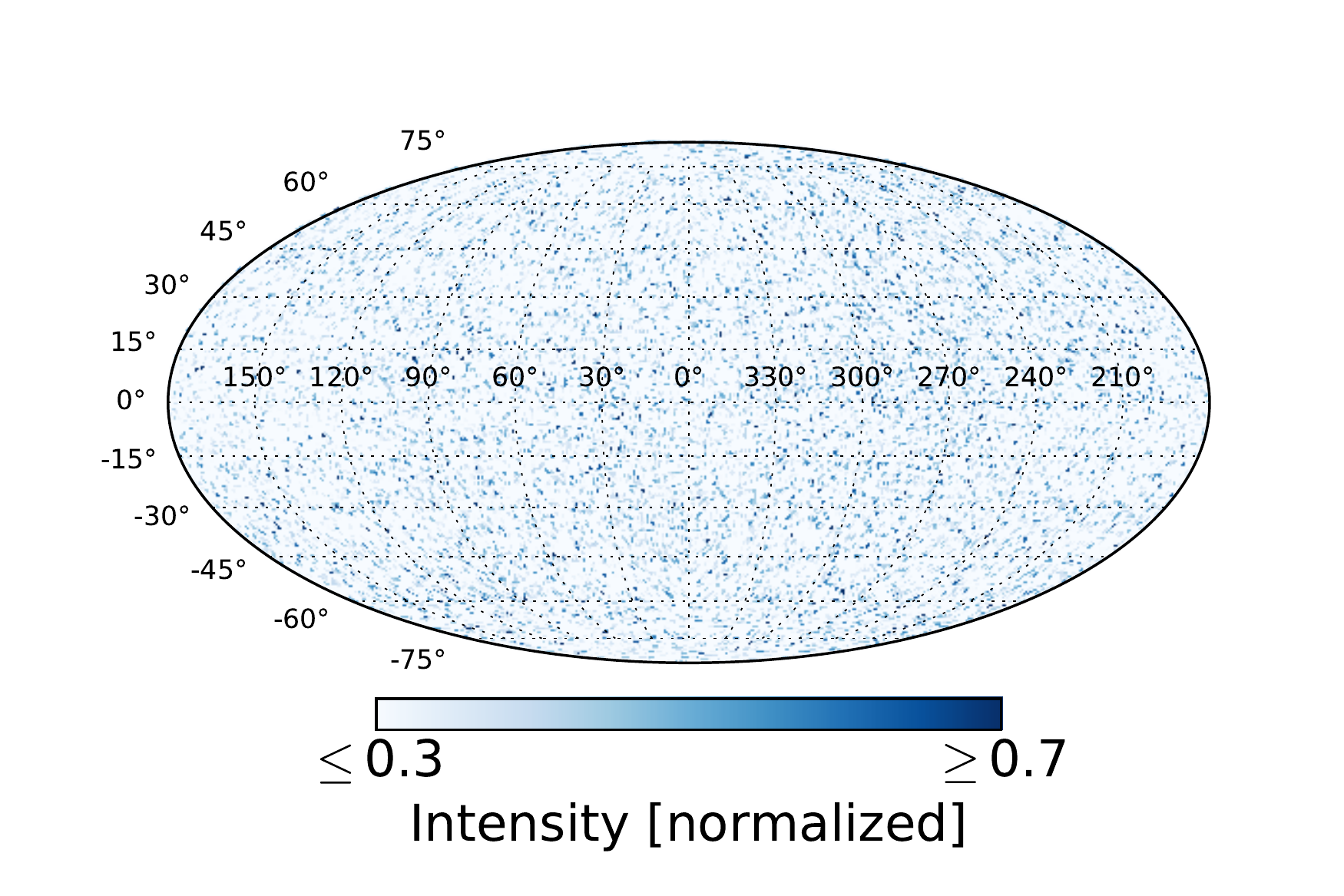}
  	}
  	\subfigure[3D proton, with galactic magnetic field]{
    	\includegraphics[width=.49\textwidth]{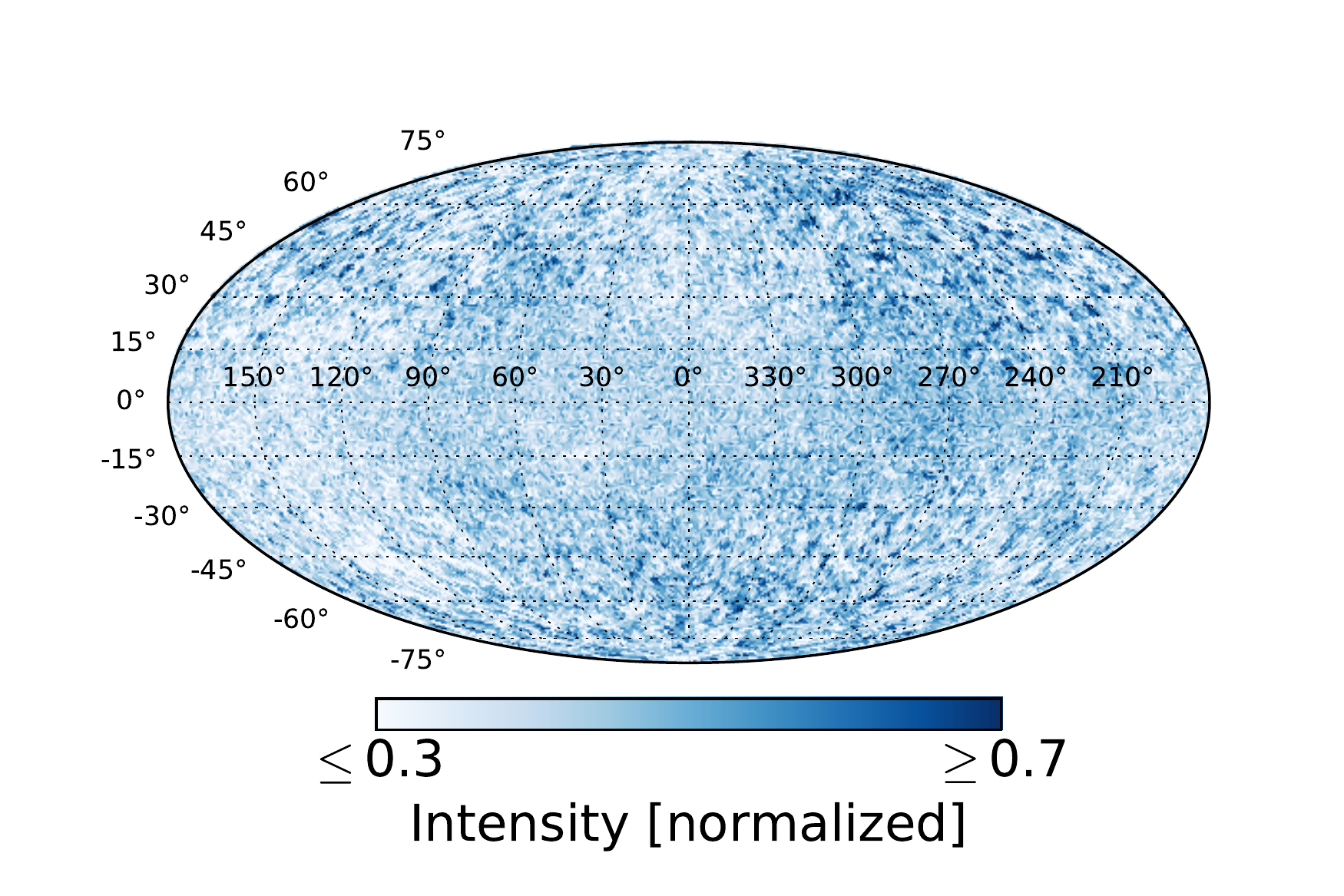}
  	}
  	\subfigure[3D iron, no galactic magnetic field]{
    	\includegraphics[width=.49\textwidth]{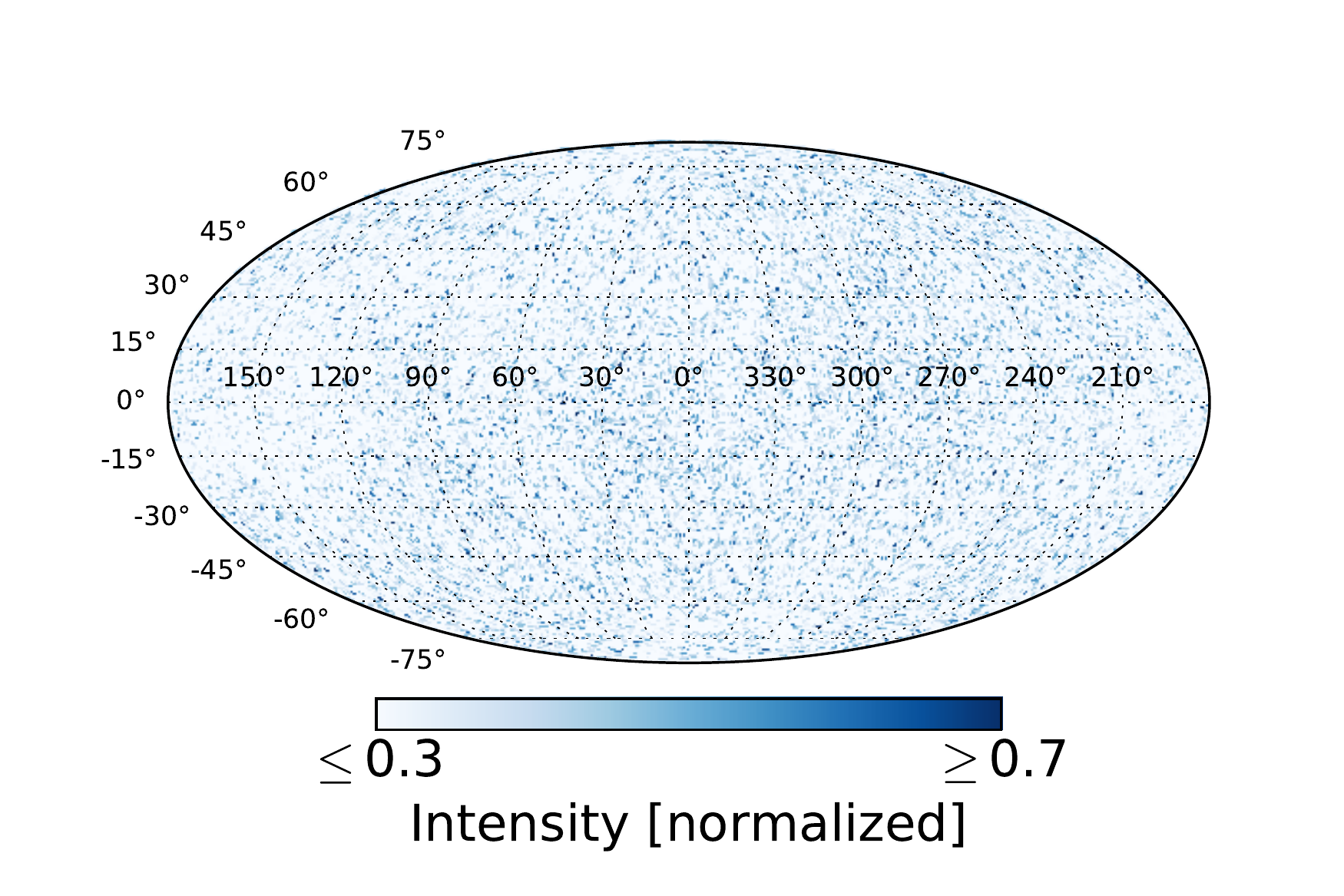}
  	}
  	\subfigure[3D iron, with galactic magnetic field]{
    	\includegraphics[width=.49\textwidth]{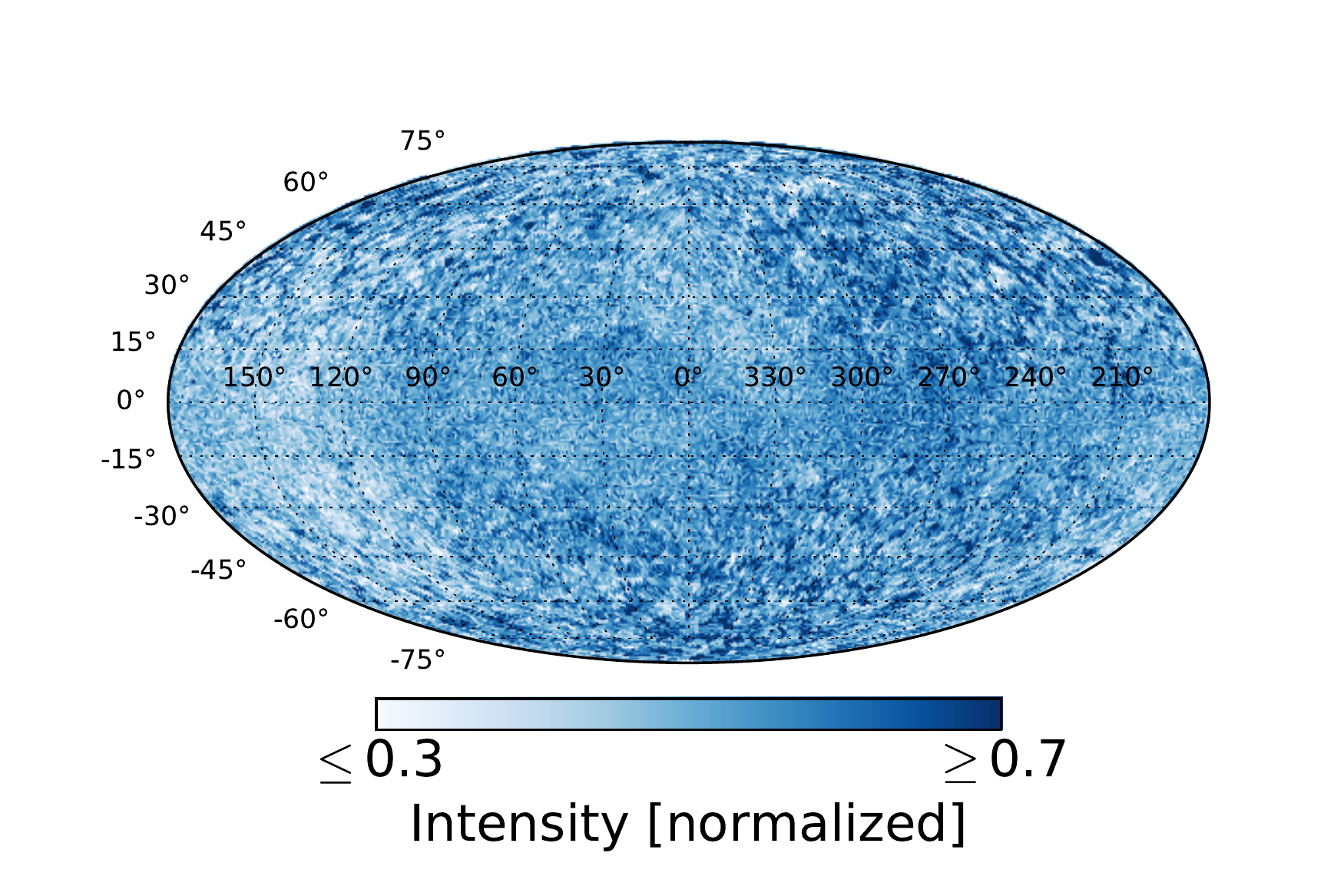}
  	}
	\caption{Sky maps in galactic coordinates for pure proton injection (top) and pure iron injection (bottom) before (left) and after (right) deflections in the galactic magnetic field are taken into account. All sky maps have been normalized so that the bin with the maximum number of hits is set to one. For better visibility we restrict the color scale to values between 0.3 and 0.7. The simulations use a spectral index of $\gamma = -1.5$ and cutoff energy $E_{\text{cut}} = 780$~EeV down to a minimum energy of $E_0 = 1$~EeV. See text for further details.}
	\label{fig:3D_skymaps}
\end{figure}

For the galactic propagation the lensing technique described in section~\ref{sec:GalProp} has been used. The implemented galactic magnetic field model is the Jansson and Farrar 2012 model~\cite{Jansson2012, Jansson2012a} including the random large-scale and turbulent small-scale component. Resulting sky maps before and after galactic magnetic field deflections for proton and iron injections at the sources are shown in figure \ref{fig:3D_skymaps}. These sky maps have been normalized so that the bin with the maximum number of hits is set to one. The range of the color scale has been set to values from 0.3 to 0.7 to better visualize the differences between the sky maps. 

\begin{figure}[tb]
	\centering
	\includegraphics[width=.6\textwidth]{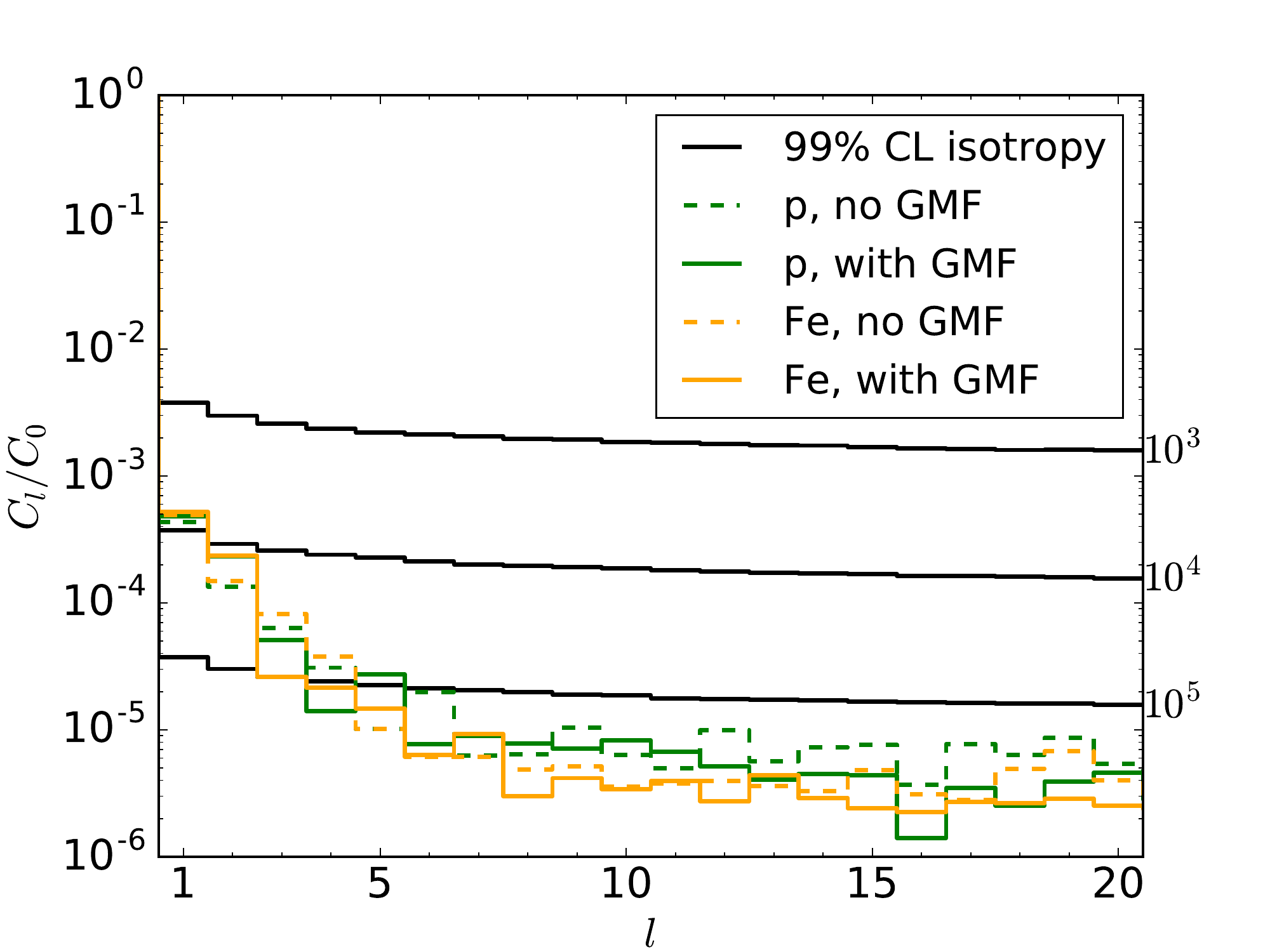}
	\caption{Angular power spectra of the cosmic ray arrival directions for pure proton (green) and pure iron (orange) injection at the sources without (dashed lines) and with (solid lines) deflections in the galactic magnetic field taken into account. Additionally, the $99\%$ confidence level upper bounds that would result from fluctuations of isotropic distributions with $10^3$, $10^4$ and $10^5$ events are shown (black solid lines). The simulations are done using a spectral index of $\gamma = -1.5$ and cutoff energy $E_{\text{cut}} = 780$~EeV down to a minimum energy of $E_0 = 1$~EeV. See text for further details.}
	\label{fig:3D_powerspectra}
\end{figure}

The coordinate-independent angular power spectra, 
\begin{equation}
	C_l = \frac{1}{2l+1}\sum^l_{m=-l}|a_{lm}|^2
\end{equation} 
with multipoles $a_{lm}$ for these sky maps are shown in figure\ \ref{fig:3D_powerspectra}, normalized to $C_0$. The sky maps and power spectra given here are for the full energy range ($\geq1$~EeV) and, as the flux is decreasing with energy, are dominated by cosmic rays with energies $\lesssim 10$~EeV. In this energy range the iron injection scenario consists mainly of light nuclei, products from photodisintegrated iron nuclei, and is therefore similar to the proton injection scenario. The dipole amplitude $r_1 = 3\sqrt(C_1/C_0)$ (see the first bin of figure\ \ref{fig:3D_powerspectra} for $C_1/C_0$) for the proton injection scenario before (after) deflections in the GMF is $r_1^{\text{p}} = 0.062$ (0.066) and for the iron injection scenario is $r_1^{\text{Fe}} = 0.067$ (0.069). For comparison the expected $99\%$ confidence level upper bounds that would result from fluctuations of isotropic distributions with $10^3$, $10^4$ and $10^5$ events are shown. This indicates the sensitivity of a hypothetical experiment with full sky coverage to detect the level of anisotropy presented in this example application for different numbers of detected events with energies above 1~EeV. The energy spectra and compositions at the observer are depicted in figures \ref{fig:4Dspec} and \ref{fig:4Dcomp} together with the spectra and compositions for the 4D cases (cf.\ section \ref{sec:4d}).

\subsection{Cosmological effects using 4D simulations}
\label{sec:4d}
As an application of the redshift dependent 4D mode we use a similar setup to the 3D example, but now including cosmological effects. Adiabatic energy losses due to the expansion of the universe are included. Events are detected when they reach the observer within a redshift window of size $\Delta z_{\rm obs} = 0.1$, i.e.\ only events that are recorded in the redshift range $-0.1 \leq z \leq 0.1$ are taken into account. For efficiency purposes we choose an observer of radius 1~Mpc, unlike in the 3D example, in which a 100~kpc observer is used. This does not significantly affect the spectrum and composition because the difference between the observer sizes in these two cases is small.

\begin{figure}[tb]
	\centering
	\includegraphics[width=0.80\columnwidth]{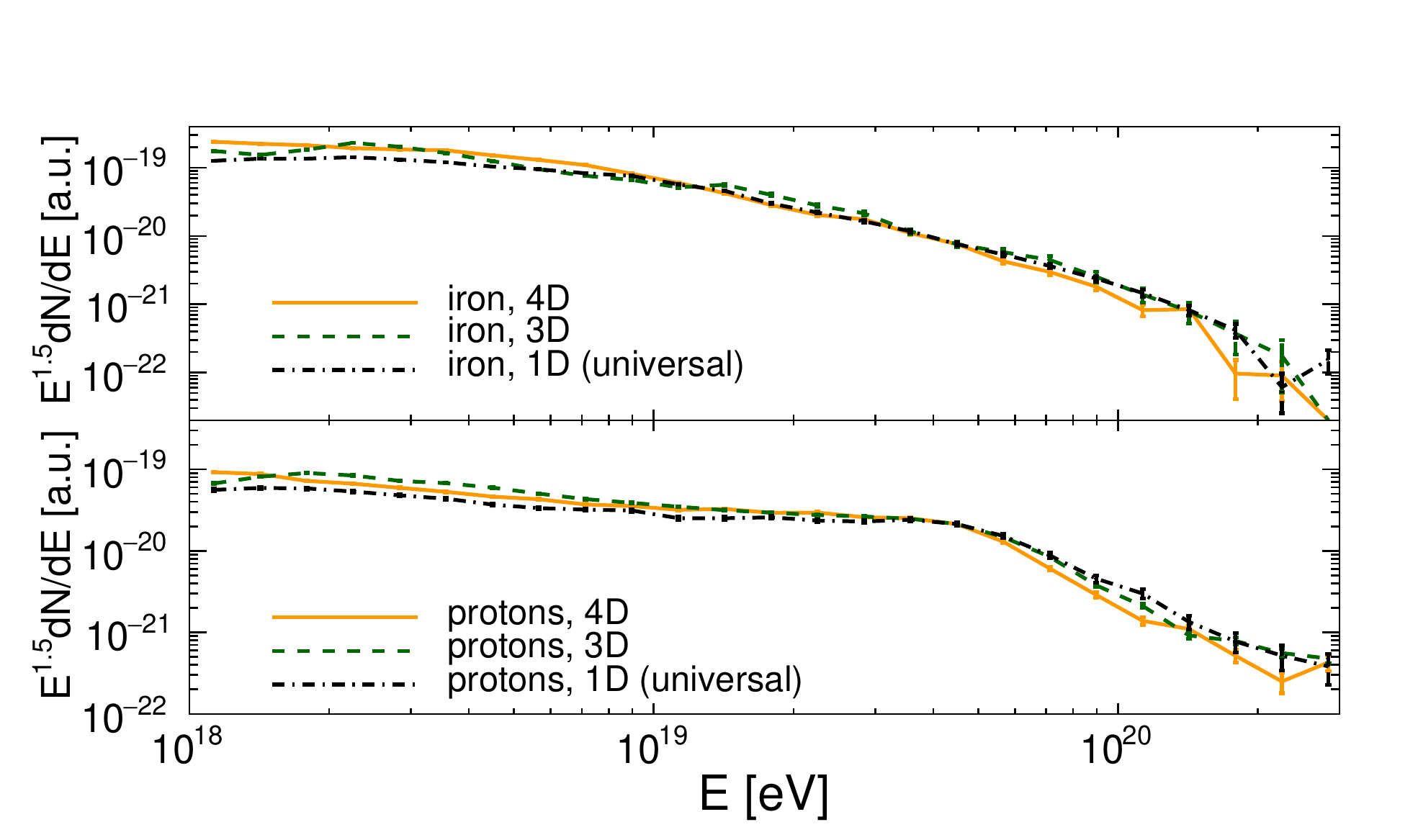}
	\caption{Simulated spectra for the proton (lower panel) and iron (upper panel) injection. The orange solid lines correspond to the 4D simulation, the 3D case is represented by green dashed lines, and dot-dashed black lines correspond to the 1D spectrum. The spectra are normalized at 10$^{20} \text{ eV}$. Parameters used in these simulation are: $\gamma=-1.5$, $E_{\rm cut}=780\text{ EeV}$, and $E_0 = 1$~EeV.}
	\label{fig:4Dspec}
\end{figure}

\begin{figure}[tb]
	\centering
	\includegraphics[width=0.49\columnwidth]{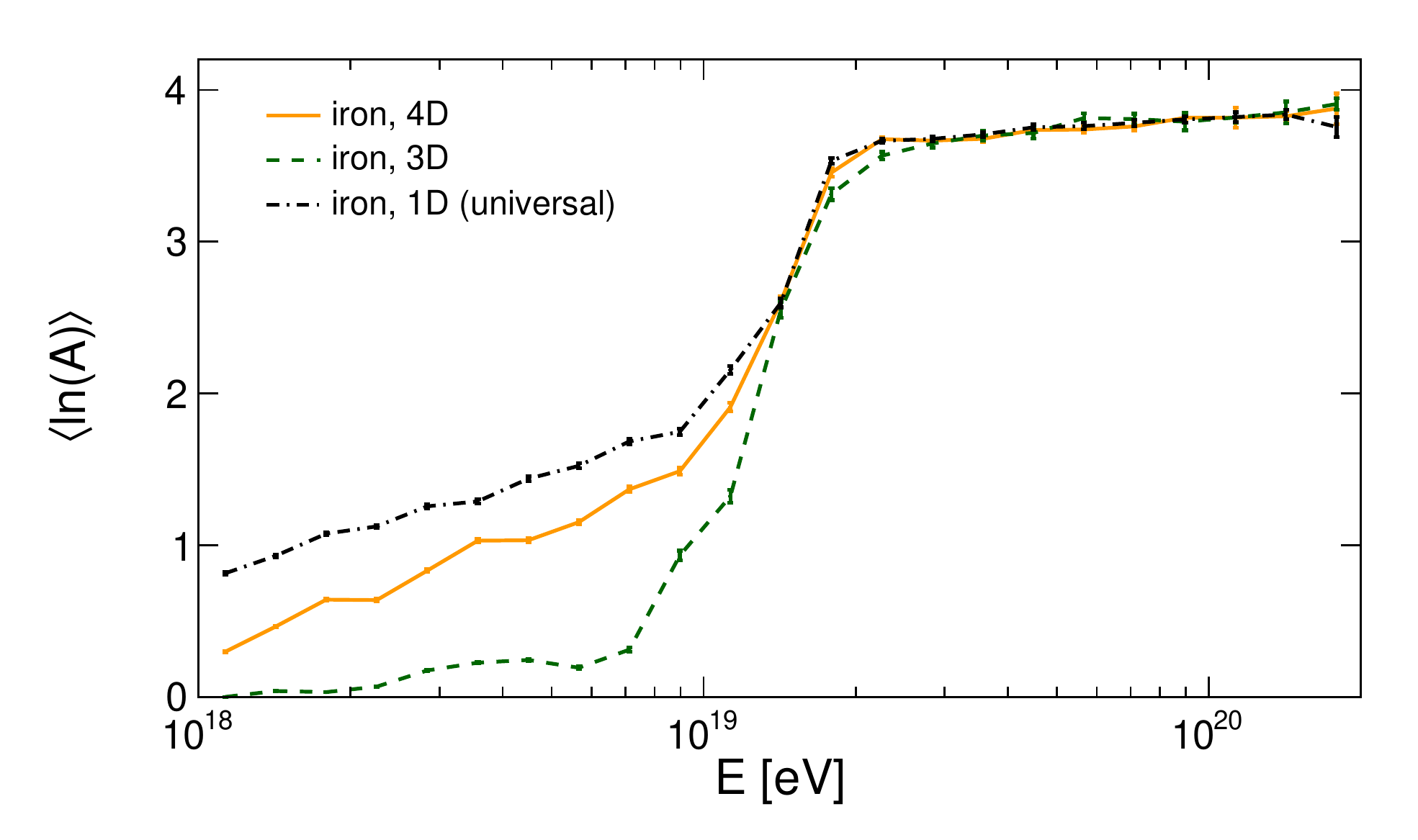}
	\includegraphics[width=0.49\columnwidth]{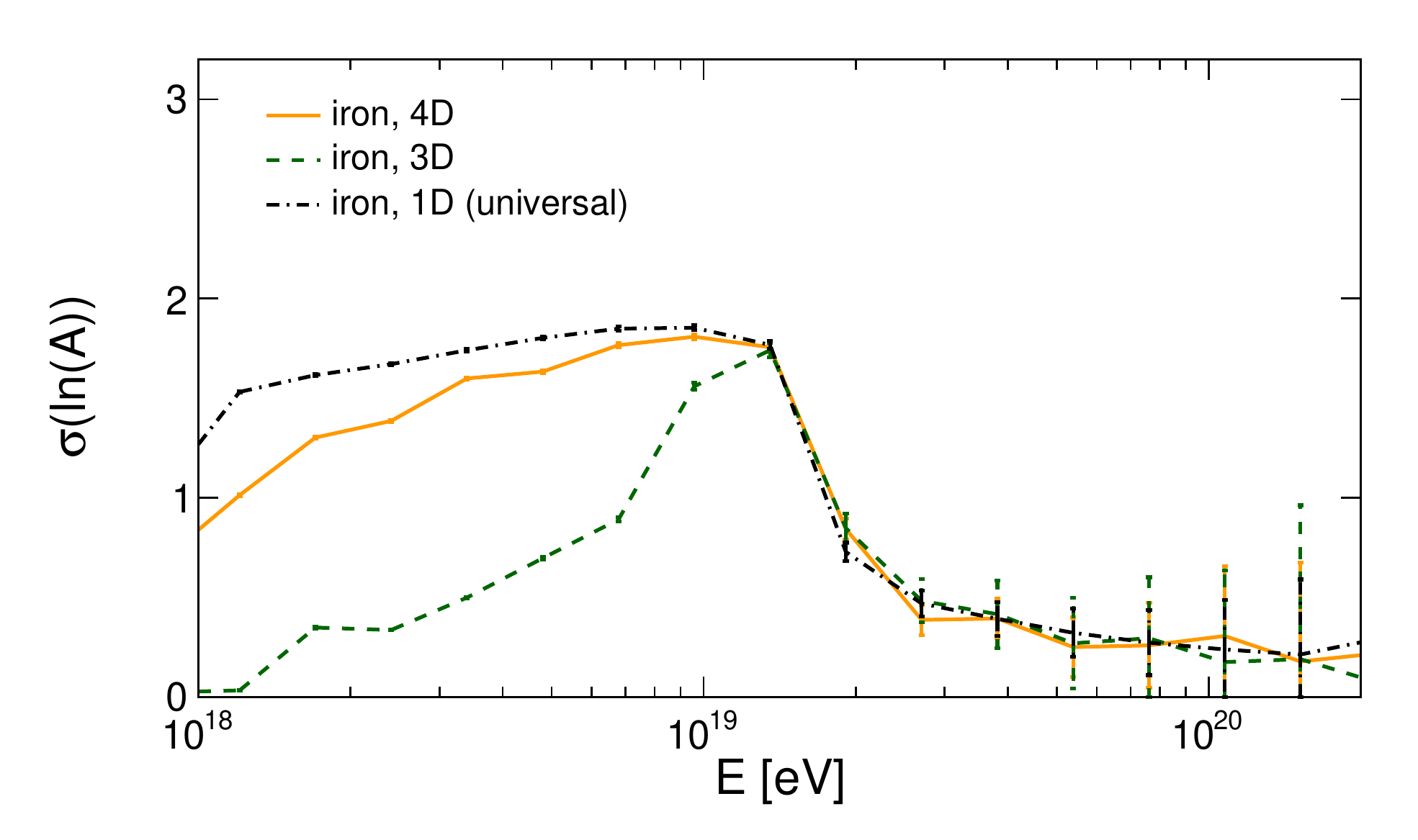}
	\caption{Mass composition observed at Earth for the case of a source emitting only iron. The average of the logarithm of the mass number, $\langle \ln A\rangle$, and its standard deviation, $\sigma(\ln A)$, are shown in the left and right panels, respectively. Orange solid lines correspond to the 4D simulation, green dashed lines correspond to the 3D case, and dot-dashed black lines represent the 1D spectrum. Parameters used in these simulation are: $\gamma=-1.5$, $E_{\rm cut}=780\text{ EeV}$, and $E_0 = 1$~EeV.}
	\label{fig:4Dcomp}
\end{figure}

We adopt the same settings as in the 3D example and simulate two scenarios: pure iron and pure proton compositions. The maximum energy and source spectral index are the same as before. The only difference is the additional energy loss process (adiabatic losses) and the redshift dependence of the CMB and IRB. The sources in the 4D scenario are assumed to have a uniform redshift distribution up to $z = 2$. The spectra for both scenarios are shown in figure~\ref{fig:4Dspec}, and the mean and variance of the logarithm of the mass composition, $\langle \ln A \rangle$ and $\sigma(\ln A)$, observed at Earth for the iron cases are shown in figure~\ref{fig:4Dcomp}, together with the universal spectrum, i.e., the spectrum obtained from a one-dimensional simulation assuming a uniform distribution of sources.  

In figure~\ref{fig:4Dspec} we notice that the differences in the spectrum are overall small, which suggests that the analyzed scenarios are fairly close to the universal case. 
Nevertheless, there are noticeable effects in the observed composition for the iron source scenario as illustrated in figure~\ref{fig:4Dcomp}. Note that the common feature of an abrupt increase in average mass at around $E \approx 10^{19.1}\,\text{eV}$ is due to kinematics: An iron nucleus injected with an energy $E_{\rm inj}$ of several hundred EeV typically suffers complete photodisintegration into 56 nucleons of energy $E_{\rm inj}/56$ each.
Due to the hard source spectrum $\propto E^{-1.5}$ these secondary protons dominate the observed flux up to the energy $780\,\text{EeV}/56 \approx 10^{19.1}\,\text{eV}$ that corresponds to the cutoff in the source spectrum of the initial iron nuclei.
One can see that for $E \lesssim 10^{19}\text{ eV}$ in the 3D case $\langle \ln A \rangle$ is smaller than in the 4D case. This is explained by the effect of adiabatic energy losses and generally stronger interactions in the 4D case, which are neglected in the 3D simulation. Indeed, additional energy losses at highest energies lower the energy up to which secondary protons can contribute and the transition is less abrupt. Similar reasoning holds for $\sigma(\ln A)$.
 
In the absence of intervening magnetic fields the composition as well as the spectrum for the 1D scenario are expected to be the same as for 4D. In this example, however, for $E \lesssim 10^{19}\text{ eV}$, the 1D scenario has a slightly larger $\langle \ln A \rangle$ with respect to 4D. This can be explained by the presence of intervening magnetic fields in combination with a discrete source distribution, which cause an increase in the effective propagation time of the particles, and hence the probability of an interaction to occur during propagation. Another effect that, although subdominant in this example, might play a role at low energies ($E \sim 1 \text{ EeV}$) is the magnetic horizon, which spawns a suppression in the flux of cosmic rays when the trajectory lengths become comparable to the age of the universe~\cite{Lemoine:2004uw,Kotera:2007ca,Mollerach:2013dza,Batista:2014xza}.

Although we have taken redshift effects into account, the comoving magnetic field can also evolve with redshift approximately as $B_{\rm com}(z) = B(0) (1+z)^{m}$, as discussed in section~\ref{Sec.CosEffects}. We have not considered the redshift evolution of the field, which is equivalent to taking $m=0$ in the aforementioned equation. The parameter $m$ is expected to be non-zero if the magnetic field is significantly affected by magnetohydrodynamic processes taking place during structure formation. Moreover, this approximation is adequate for very high energies ($E \gtrsim 10^{19} \text{ eV}$), where cosmic rays most likely originate in the nearby universe. At energies of the order of a few EeV the redshift evolution of the magnetic field might play an important role. We note, however, that even though the redshift evolution of the magnetic field can be taken into account in CRPropa this is outside the scope of the present work.

\section{Summary}
\label{Sec:Summary}
The simulation of galactic and extragalactic cosmic ray propagation plays an essential role in understanding astrophysical processes at ultra-high energies. In this paper we introduced the new version of the publicly available cosmic ray propagation code CRPropa 3. To interpret the data collected by large-scale cosmic ray observatories, the code was completely rewritten to be flexible enough to cover the large parameter space of astrophysical scenarios. The modular structure, included in version 3, enables to combine independent modules to study multiple use cases and even to extend the code by new individually specified modules.

While inheriting all features from the previous version, CRPropa 3 introduces additional functionalities. As a consequence of the modular and flexible code structure any kind of magnetic field is now supported and additionally modelization of deflection in galactic magnetic fields has been improved.
Furthermore, the lensing technique provides a computationally efficient method to calculate trajectories of cosmic rays inside the Milky Way and can be applied for the most commonly used galactic magnetic field models. 

The calculation of photonuclear interactions has been updated to latest models and a number of new IRB models have been implemented. In the case of 3D simulations cosmological effects are now taken into account and the evolution of the IRB with redshift is now implemented with a redshift dependent scaling. To take into account the \textit{a priori} unknown propagation length and the resulting changes in photon background, CRPropa 3 extends the notion of a 3D observer to a 4D mode by taking the redshift (time) into account. 

To complete the multi-messenger picture, CRPropa 3 includes the production and propagation of secondary particles such as neutrinos and gamma rays. In addition to the transport code DINT for electromagnetic cascades we now incorporated a full Monte Carlo code to calculate the electromagnetic cascade at energies above 0.1~EeV based on the EleCa code. A combination of the two codes is available enabling to follow cascades down to $\sim 100$~TeV.

Information on downloading the code, usage and example applications can be found at \url{http://github.com/CRPropa/CRPropa3}. Questions and comments can be submitted to the ticketing system \url{https://github.com/CRPropa/CRPropa3/issues}. The features of CRPropa 3 can also be explored online without installation at \url{https://vispa.physik.rwth-aachen.de}.

\appendix
\section{Giant dipole resonance parameters}
\label{appendix:GDR}
\begin{figure}
\centering
\includegraphics[width=0.495\textwidth]{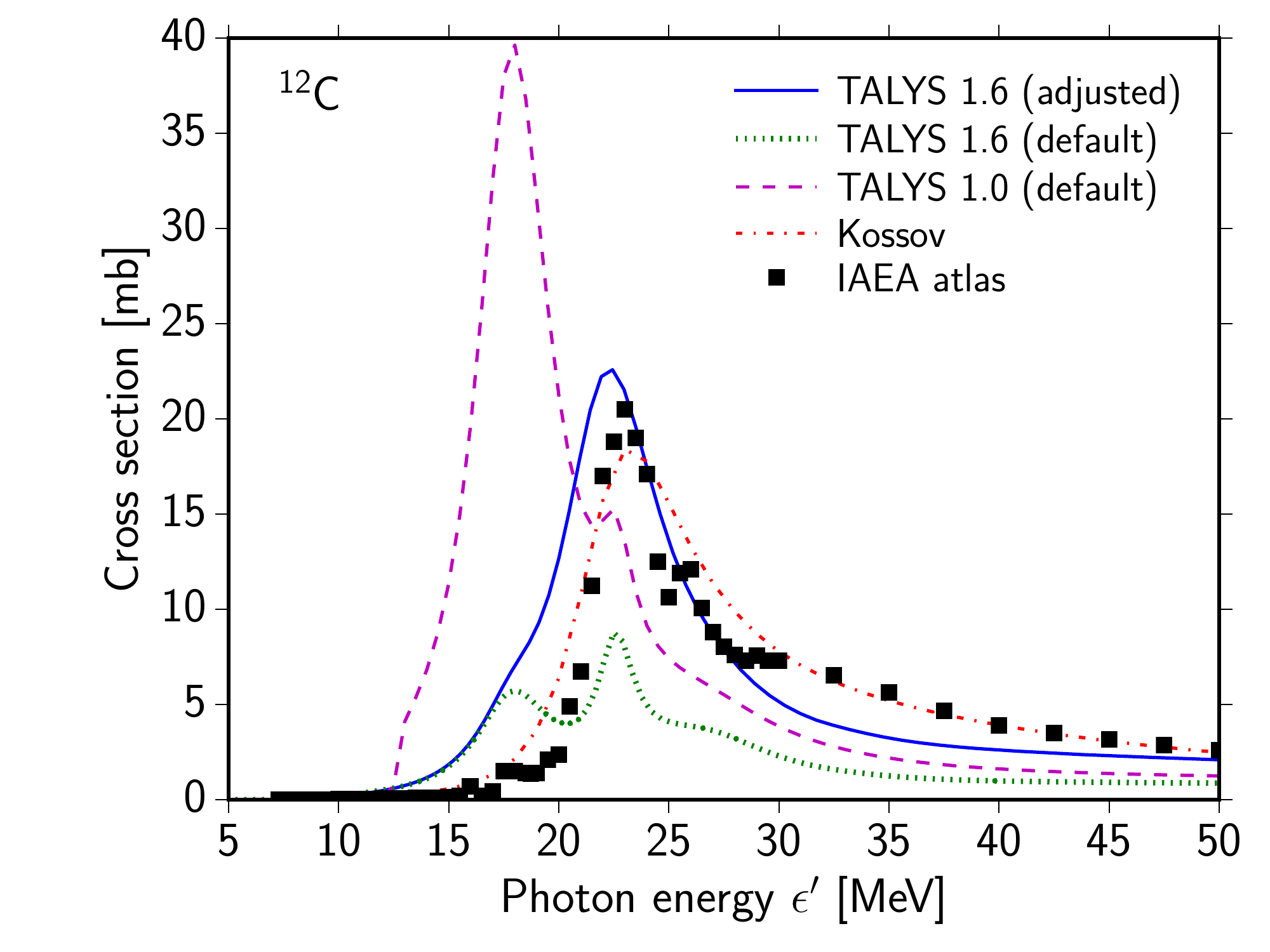}
\includegraphics[width=0.495\textwidth]{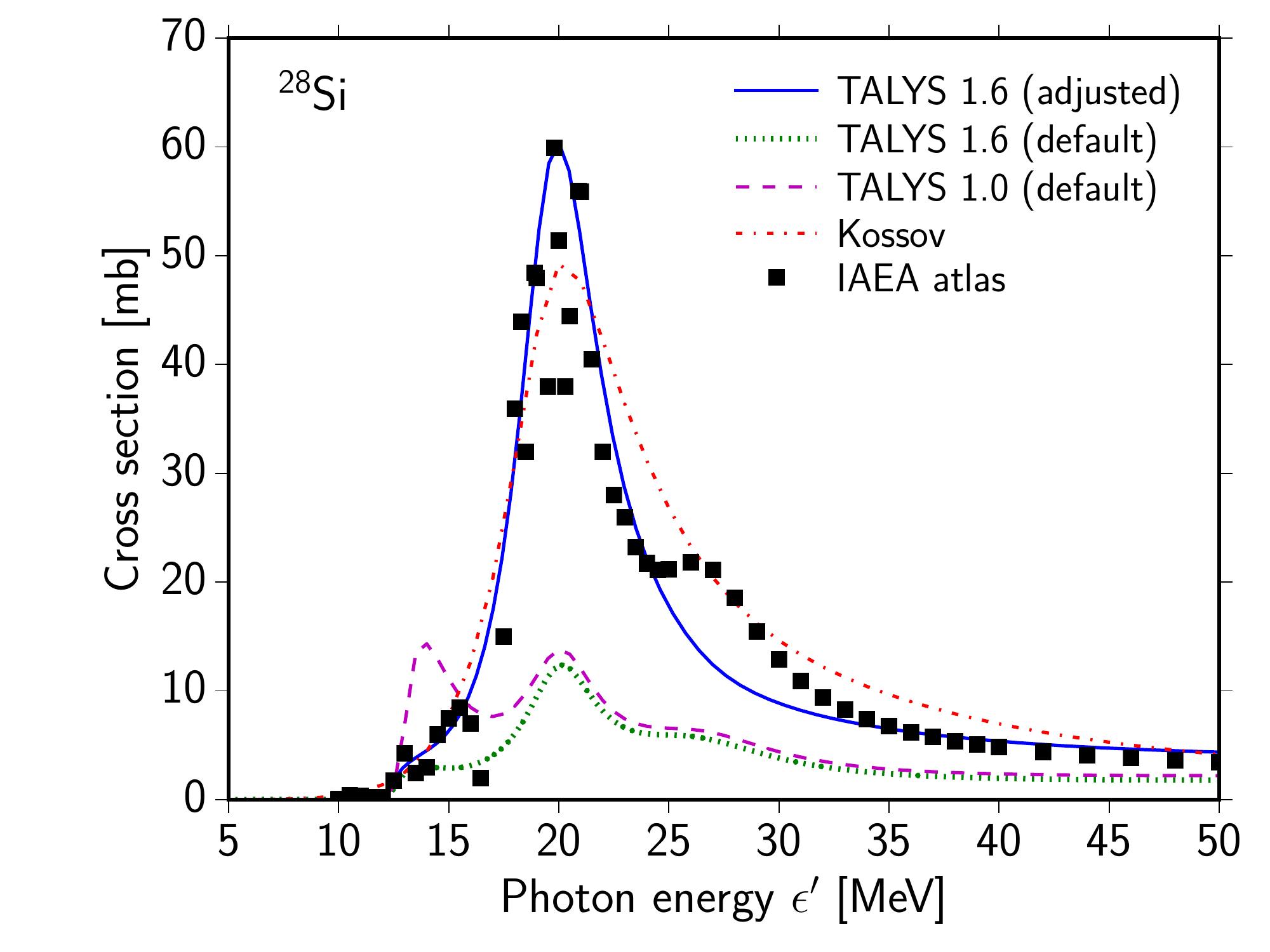}
\caption{Comparison of total photodisintegration cross sections for $^{12}$C (left) and $^{28}$Si (right) with the evaluated experimental data compiled in the IAEA atlas~\cite{iaea-tecdoc-1178}. TALYS 1.0 (default) and TALYS 1.6 (adjusted) correspond to models implemented in CRPropa~2 and CRPropa~3, respectively. Alternatively, the Kossov parametrization~\cite{Kossov2002} can be used.}
\label{fig:C12}
\end{figure}

Photodisintegration is the most important interaction for cosmic ray nuclei with energies $E > 10^{19}$\,eV.
Cross sections for this interaction are dominated by the giant dipole resonance for photons with energies $\epsilon' < 30$\,MeV in the nucleus rest frame.
In ref.~\cite{Khan:2004nd} an ``accurate description'' of the available experimental data was found using a preliminary version of TALYS.
TALYS was used in this comparison~\cite{private_communication_talys} with the giant dipole resonance parameters from the IAEA atlas~\cite{iaea-tecdoc-1178}.
In contrast, the publicly available versions of TALYS by default uses the giant dipole resonance parameters from the RIPL-2 database~\cite{iaea-tecdoc-1506},
and the predicted cross sections are in poor agreement with the experimental data.
Thus, in CRPropa~3 TALYS is used with the giant dipole resonance parameters of the IAEA atlas, if available, as the resulting cross sections are in much better agreement with the available measurements.
In figure~\ref{fig:C12} we show the total photodisintegration cross sections for $^{12}$C and $^{28}$Si. The Kossov~\cite{Kossov2002} parametrization and TALYS 1.6 with adjusted giant dipole resonance parameters are in reasonable agreement with experimental data. Both models are implemented in CRPropa~3.
The complete list of used giant dipole resonance parameters is given in Table~\ref{tab:GDRparameters}.

\begin{table}[t]
\centering
\begin{tabular}{| l || c | c | c | c | c | c | c |}
    \hline
    Isotope & $E_0$\,[MeV] & $\sigma_0$\,[mb] & $\Gamma_0$\,[MeV] & $E_1$\,[MeV] & $\sigma_1$\,[mb] & $\Gamma_1$\,[MeV] & Source \\
    \hline
    $^{12}$C   &  22.70  &  21.36  &  6.00  &  &  &  &  Atlas \\
    $^{14}$N   &  22.50  &  27.00  &  7.00  &  &  &  &  Atlas \\
    $^{16}$O   &  22.35  &  30.91  &  6.00  &  &  &  &  Atlas \\
    $^{23}$Na  &  23.00  &  15.00  & 16.00  &  &  &  &  Atlas \\
    $^{24}$Mg  &  20.80  &  41.60  &  9.00  &  &  &  &  Atlas \\
    $^{27}$Al  &  21.10  &  12.50  &  6.10  &  29.50  &  6.70  &  8.70 &  RIPL-2 \\
    $^{28}$Si  &  20.24  &  58.73  &  5.00  &  &  &  &  Atlas \\
    $^{40}$Ar  &  20.90  &  50.00  & 10.00  &  &  &  &  Atlas \\
    $^{40}$Ca  &  19.77  &  97.06  &  5.00  &  &  &  &  Atlas \\
    $^{51}$V   &  17.93  &  53.30  &  3.62  &  20.95  &  40.70  &  7.15 &  RIPL-2 \\
    $^{55}$Mn  &  16.82  &  51.40  &  4.33  &  20.09  &  45.20  &  4.09 &  RIPL-2 \\
    \hline
\end{tabular}
\caption{Giant dipole resonance parameters used with TALYS (as parameters for the Kopecky-Uhl generalized Lorentzian model of the E$1$-strength function): peak energy $E_i$, peak cross section $\sigma_i$ and width $\Gamma_i$ for resonances with a single ($i=0$) or a split peak ($i=0,1$). Default values from the RIPL-2 database~~\cite{iaea-tecdoc-1506} are replaced, if available, with the total cross section parameters from the atlas of giant dipole resonance parameters. Note that for isotopes not listed, as well as for higher order contributions, TALYS uses a compilation of formulas listed in \cite{TALYSmanual}.}
\label{tab:GDRparameters}
\end{table}

\acknowledgments
The authors would like to thank Carmelo Evoli for contributions to the code, Lukas Merten for helpful comments and suggestions and Christopher Heiter and Mariangela Settimo for improvements in the photon propagation calculation. Thanks also to Marcus Wirtz for improving the galactic magnetic field modeling. We are grateful to Arian Koning and St\'ephane Goriely for guidance in using the TALYS package. This work was supported by the Deutsche Forschungsgemeinschaft (DFG) through the collaborative research centre SFB 676, by the Bundesministerium f\"ur Bildung und Forschung (BMBF) and by the Helmholtz Alliance for Astroparticle Physics (HAP) funded by the Initiative and Networking Fund of the Helmholtz Association including the cost takeover of the open access processing charge. AvV acknowledges financial support from the NWO astroparticle physics grant WARP.


\end{document}